\documentclass[10pt]{article}
\usepackage{epsfig,psfig,graphicx,rotating}
\def\be{\begin{equation}}
\def\ee{\end{equation}}
\newcommand{\bea}{\begin{eqnarray}}
\newcommand{\eea}{\end{eqnarray}}
\begin{document}
\title{\bf  
STATISTICAL MECHANICS OF THE SELF-GRAVITATING GAS: THERMODYNAMIC
LIMIT, PHASE DIAGRAMS AND FRACTAL STRUCTURES\footnote{Lectures given at the  
7th. Paris Cosmology  Colloquium, Observatoire de Paris, June 11-15, 2002,
Proceedings edited by H. J. de Vega and N. G. S\'anchez and published by 
Observatoire de Paris (2004), and at the 9th Course of the International School of 
Astrophysics `Daniel Chalonge', Palermo, Italy, 
7-18 September 2002, NATO ASI, Proceedings
edited by N. G. S\'anchez and Y. Parijskij, Kluwer, Series II, vol. 130, 2003.}}
\author{{\bf  H. J. de Vega}(a,b)\footnote{devega@lpthe.jussieu.fr}, 
{\bf N. G. S\'anchez}(b)\footnote{Norma.Sanchez@obspm.fr} \\ \\
(a)Laboratoire de Physique Th\'eorique et Hautes Energies, \\
Universit\'e Paris VI et Paris VII, 
Laboratoire Associ\'e au CNRS UMR 7589. \\ 
Tour 24-25, 5\`eme \'etage, 
4, Place Jussieu 75252 Paris, Cedex 05, France. \\
(b) Observatoire de Paris,  LERMA, 
Laboratoire Associ\'e au CNRS UMR 8112, 
\\ 61, Avenue de l'Observatoire, 
75014 Paris,  France.}
\maketitle
\begin{abstract}
We provide a complete picture of the non-relativistic self-gravitating 
gas at thermal equilibrium using Monte Carlo simulations,
analytic mean field methods (MF) and low density expansions. The
system is shown to possess an infinite volume limit in the grand
canonical (GCE), canonical (CE) and microcanonical (MCE) ensembles when
$(N, V)  \to \infty$, keeping 
$N/ V^{1/3}$ fixed. We {\bf compute} the equation of state (we do not assume
it as is customary), as well as the energy, free energy, entropy, 
chemical potential, specific heats, compressibilities and speed of
sound; we analyze their properties, signs and singularities.  All
physical quantities turn out to depend on a single variable $ \eta
\equiv {G \, m^2 N \over V^{1/3} \; T}$ that is kept fixed in the $
N\to \infty $ and $ V \to 
\infty $ limit. The system is in a  gaseous phase for $ \eta < \eta_T $ and
collapses into a dense object for $ \eta > \eta_T $ in the CE with the pressure
becoming large and negative. At $ \eta = \eta_T $ the isothermal
compressibility diverges; this gravitational  phase transition is
associated to the Jeans' instability, our Monte Carlo simulations yield
$ \eta_T \simeq 1.515 . \; PV/[NT] = f(\eta) $ and all physical magnitudes 
exhibit a square root branch point at $ \eta = \eta_C >  \eta_T $.
The values of  $ \eta_T $ and $ \eta_C $ change by a few percent
with the geometry 
for large $ N $: for spherical symmetry and $ N = \infty $ (MF), we
find $ \eta_C = 1.561764\ldots $,  while the Monte Carlo simulations for
cubic geometry yields $ \eta_C \simeq 1.540 $. In mean field and
spherical symmetry $c_V$ diverges as $
\left(\eta_C-\eta \right)^{-1/2} $ for $ \eta \uparrow
\eta_C $ while $ c_P $ and $ \kappa_T $ diverge as $ \left(\eta_T-\eta
\right)^{-1} $ for $ \eta \uparrow \eta_T = 1.51024\ldots  $.
The function $ f(\eta) $ has a
second Riemann sheet which is only physically realized in the MCE. In the MCE, 
the collapse phase transition takes place in this second sheet near $
\eta_{MC} = 1.26 $ and the pressure and temperature are
larger in the collapsed phase than in the gaseous phase. Both
collapse phase transitions (in the CE and in the MCE) are of zeroth
order since the Gibbs free energy has a jump at the transitions. 
The MF equation of state in a sphere, $ f(\eta)$, obeys a {\bf first
order} non-linear differential equation of first kind Abel's type.  
The MF gives an extremely  accurate picture in agreement with the MC
simulations both in the CE and MCE. We perform the MC
simulations on a cubic geometry, which thus describe an {\bf isothermal cube}
while the MF calculations describe an isothermal sphere.

We complete our study of the self-gravitating gas by computing
the fluctuations around the saddle point solution for the three
statistical ensembles (grand canonical, canonical and
microcanonical).  Although the saddle point is the same for the three
ensembles, the fluctuations change from one ensemble to the other. The
zeroes of the small fluctuations determinant determine the  position
of the critical points for each  ensemble. This yields the domains of
validity of the mean field approach.  Only the S wave
determinant exhibits critical points. Closed formulae for
the S and P wave determinants of fluctuations are derived.
The {\bf local} properties of the self-gravitating gas in thermodynamic
equilibrium are studied in  detail.
The pressure, energy density, particle density and speed
of sound are computed and analyzed as functions of the position. 
The equation of state turns out to be {\bf
locally } $ p(\vec r) = T \, \rho_V(\vec r) $ as for the ideal gas.  
Starting from the partition function of the self-gravitating gas,
we prove in this microscopic calculation that
the hydrostatic description yielding locally the ideal  gas equation of
state is exact in the $ N = \infty $ limit. The dilute nature of the
thermodynamic limit ($N \sim L \to \infty $ with $N/L$ fixed) together
with the long range nature of the gravitational forces
play a crucial role in obtaining such ideal gas equation.
The self-gravitating gas being inhomogeneous, we have   $ PV/[NT] = f(\eta)
\leq  1 $ for any finite volume $V$.  The
inhomogeneous particle distribution in the ground state suggests a
fractal distribution with Haussdorf dimension $D$, $D$ is slowly decreasing
with increasing density, $ 1 < D < 3$.
The average distance between particles is computed in Monte Carlo
simulations and analytically in the mean field approach. A dramatic
drop  at the phase transition is exhibited, clearly
illustrating the properties of the collapse.
\end{abstract}
\tableofcontents
\section{Statistical Mechanics of  the Self-Gravitating  Gas}

Physical systems at thermal equilibrium are usually homogeneous. This is the 
case for gases with short range  intermolecular forces (and in absence
of external fields). In such cases the entropy is maximum when the
system homogenizes.

When long range interactions as the gravitational force are present, even
the ground state is inhomogeneous. In this case,  each element of the
substance is acted on by very strong forces due to distant
particles in the gas. Hence, regions near to and far from the boundary of the 
volume occupied by the gas will be in very different conditions, and, as a 
result, the homogeneity of the gas is destroyed \cite{llms}. The state
of maximal entropy for gravitational systems is {\bf inhomogeneous}. 
This  basic  inhomogeneity suggested us that fractal
structures can arise in a self-interacting gravitational
gas \cite{I,II,natu,sieb}. 

The inhomogeneous character of the ground state for gravitational
systems explains why the universe is {\bf not} going towards a `thermal
death'. A `thermal death' would mean that the universe evolves towards
more and more homogeneity. This can only happen if the entropy is
maximal for an homogeneous state. Instead, it is the opposite what
happens, structures are formed in the universe through
the action of the gravitational forces as time evolves.

Usual theorems in statistical mechanics break down for inhomogeneous
ground states. For example, the specific heat may be negative in the
microcanonical ensemble (not in the canonical ensemble where it is
always positive)\cite{llms}. 

As is known, the thermodynamic limit for self-gravitating systems does
not exist in its usual form ($N\to \infty,\; V \to \infty,\; N/V = $
fixed). The system collapses into a very dense phase which is
determined by the short distance (non-gravitational) forces between
the particles. However, the thermodynamic functions {\bf exist} in the 
{\bf dilute} limit \cite{I,II}
$$ 
N\to \infty\; ,\; V \to \infty\; ,\; {N\over V^{1/3}} = \mbox{fixed} \; ,
$$ 
where $ V $ stands for the volume of the box containing the gas.
In such a limit, the energy $E$, the free energy and the entropy turns to be
extensive. That is, we find that they take the form of $ N $ times a
function of the intensive dimensionless variables:
$$
\eta = {G \, m^2 N \over L \; T} \quad \mbox{or} \quad
\xi = { E \, L \over G \, m^2 \, N^2}
$$
where $\eta$ and $\xi$ are  intensive variables. Namely, $\eta$ and
$\xi$ stay finite when $ N $ and $ V \equiv L^3 $ tend to infinite. 
The variable $\eta$ is
appropriate for the canonical ensemble and $\xi$ for the
microcanonical ensemble. Physical magnitudes as the specific heat,
speed of sound, chemical potential and  compressibility only depend on
$\eta$ or $\xi$. The variables $\eta$ and $\xi$, as well as the ratio 
$ N/L $, are therefore  {\bf intensive} magnitudes. 
The energy, the free energy, the 
Gibbs free energy and the entropy are of the form $ N $ times a
function of $\eta$. These functions of $\eta$ have a finite $ N = \infty $
limit for fixed $\eta$ (once the ideal gas contributions are
subtracted). Moreover, the dependence on $ \eta $ in all these
magnitudes express through a single universal function $ f(\eta) $. 
The variable $\eta$ is the ratio of the characteristic gravitational energy
$\frac{G m^2 N}{L}$ and the kinetic energy $T$ of a particle in the gas. 
For $\eta=0$ the ideal gas is recovered.

\begin{figure}
\begin{turn}{-90}
\epsfig{file=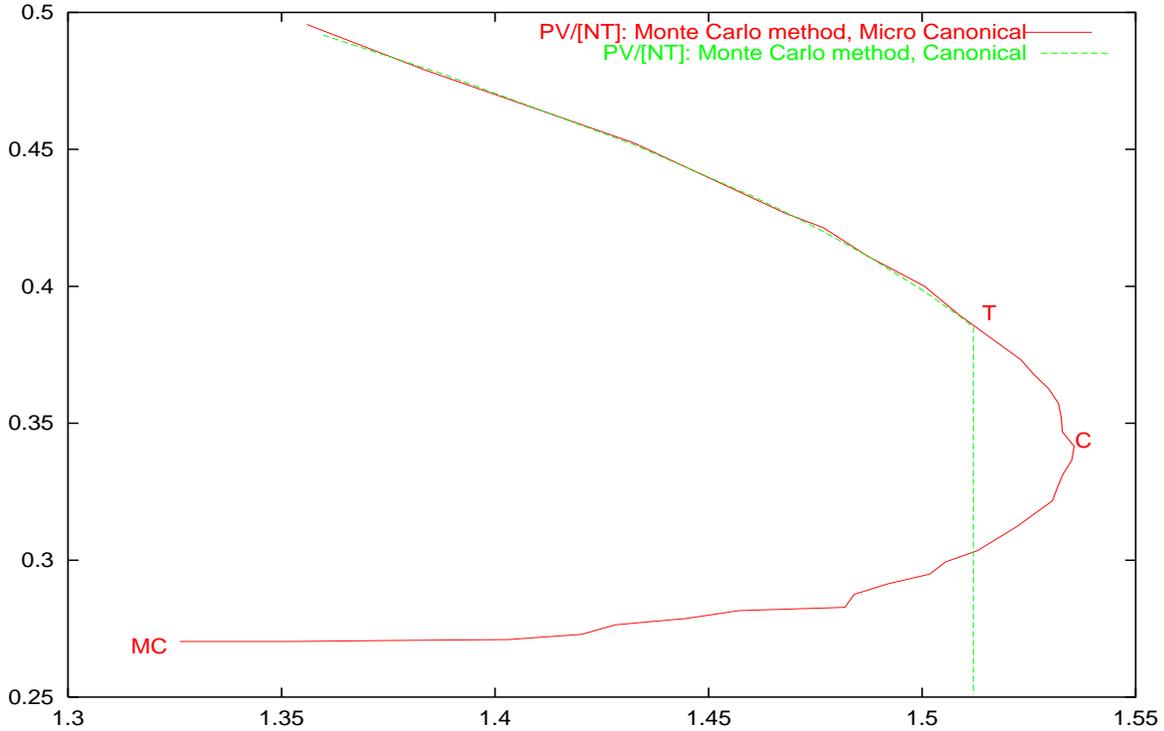,width=10cm,height=16cm} 
\end{turn}
\caption{ $ f(\eta^R) = P V/[ N T] $ as a function of $ \eta^R $  by Monte
Carlo simulations for the microcanonical and canonical ensembles
($N=2000$). Both curves coincide within the statistical error till the point T.
\label{fig14}}
\end{figure}

In refs. \cite{I,II} we have thoroughly studied
the statistical mechanics of the self-gravitating
gas. That is,  our starting point is the partition function for
non-relativistic particles interacting through their gravitational
attraction in thermal equilibrium. We study the self-gravitating gas in the
three ensembles: microcanonical (MCE), canonical (CE) and grand
canonical (GCE). We performed calculations by three methods:

\begin{itemize}

\item{By expanding
the partition function through direct calculation in powers of $1/\xi$
and $\eta$ for the MCE and CE, respectively.  These expressions apply
in the dilute regime ($ \xi \gg 1 \, , \, \eta \ll 1 $) and become
identical for both ensembles for $ N \to \infty $. At $ \eta = 0 1/\xi$ 
we recover the ideal gas behaviour.}

\item{By performing Monte Carlo simulations both in the MCE
and in the CE.  
We found in this way that the self-gravitating gas {\bf collapses} at a
critical point which depends on the ensemble considered. As shown in
fig. \ref{fig14} the collapse occurs first in the canonical ensemble (point
T). The microcanonical ensemble exhibits a larger region of stability
that ends at the point MC (fig.  \ref{fig14}). Notice that the
physical magnitudes are identical in the common region of validity of
both ensembles within the statistical error. Beyond the critical point
T the system 
becomes suddenly extremely compact with a large negative pressure in
the CE. Beyond the point MC in the MCE the pressure and the
temperature increase suddenly and the gas collapses. 
The {\bf phase transitions} at T and at MC are of {\bf zeroth order} since
the Gibbs free energy has discontinuities in both cases.}

\item{By using the mean field  approach we evaluate the partition
function for large $ N $. We do this computation  in the grand canonical,
canonical and microcanonical ensembles. In the three cases, the
partition function is expressed as a functional integral over a
statistical weight which depends on the (continuous) particle
density. These statistical weights are of the form of the exponential
of an `effective action' proportional to $ N $. Therefore, the $ N \to
\infty $ limit follows by the saddle point method. The saddle point is
a space dependent mean field showing the inhomogeneous character of the ground
state. Corrections to the mean field are of the order $ 1/N $ and can
be safely ignored for $ N \gg 1 $ except near the critical
points. These mean field results turned out to be in  excellent agreement
with the Monte Carlo results and with the low density expansion. }
\end{itemize}

We calculate the saddle point (mean field)  for spherical symmetry
and we obtain from it the various physical
magnitudes (pressure, energy, entropy, free energy, specific heats,
compressibilities, speed of sound and particle density).
Furthermore, we computed in ref.\cite{II} the {\bf determinants of small
fluctuations} around the saddle point solution for spherical symmetry
for the three statistical ensembles.

When any small fluctuation around the saddle point decreases the
statistical weight in the functional integral, the saddle point 
dominates the functional integral and the mean field approach can be valid. 
In that case, the  determinant of small fluctuations is positive. A
negative determinant of small fluctuations indicates that some
fluctuations around the saddle point are increasing the statistical
weight in the functional integral and hence the saddle point {\bf does
not} dominate the partition function. The mean field approach cannot be
used when the determinant of small fluctuations is negative. 
We find analytically
in the CE that the determinant of small fluctuations vanishes at the point 
$\eta_C$ = 1.561764 and becames negative at $\eta > \eta_C$ \cite{I}. 
(The point $\eta_C$ is indicated $C$ in fig.\ref{fig14}). 
We find that the CE specific heat $c_V$ at the point $C$ diverges as  \cite{I}:
$$
c_V \buildrel{ \eta \uparrow \eta_C}\over= \pm 0.63572\ldots
(\eta_C-\eta)^{-1/2} - 0.19924\ldots+ {O}(\sqrt{\eta_C-\eta})
$$
The (+) sign refers to the positive (first) branch, and the (-) 
sign to the negative (second) branch (between the points $C$ and $MC$). 

However, it must be noticed that the {\bf instability point} is located 
at $\eta =\eta_T < \eta_C$, as shown by both, mean field and Monte Carlo 
computations. (The point $ \eta_T $ is indicated $ T $ in fig. \ref{fig14}).
The onset of instability in the canonical ensemble
coincides with the point where the isothermal compressibility diverges. 
The isothermal compressibility $\kappa_T$ is positive from $\eta=0$ 
till $ \eta=\eta_T= 1.51024 \ldots $. At this point $ \kappa_T $ as well
as the specific heat at constant pressure $ c_P $ diverge and change
their signs \cite{I}. Moreover, at this point 
the speed of sound at the center of the sphere becomes imaginary \cite{II}. 
Therefore, small density fluctuations will {\bf grow} exponentially 
in time instead of exhibiting 
oscillatory propagation. Such a behaviour leads to the collapse of the gas into
a extremely compact object.
Monte-Carlo simulations confirm the presence of this instability at 
$ \eta_T = 1.510 \ldots$ in the canonical ensemble and the formation
of the collapsed object \cite{I}. 

The collapse in the Grand Canonical ensemble (GCE) occurs for a smaller 
value of $ \eta = \eta_{GC} = 0.49465\ldots $, while in the microcanonical
ensemble, the collapse arrives later, in the second sheet, at 
$ \eta =  \eta_{MC} = 1.25984\ldots$.

We find that the Monte Carlo simulations for self-gravitating gas in
the CE and the MCE confirm the stability results obtained from mean field. 

The saddle point solution is {\bf identical} for the three statistical
ensembles. This is not the case for the fluctuations around it. The
presence of constraints in the CE (on the number of particles) and in
the MCE (on the energy and the number of particles) {\bf changes} the
functional integral over the quadratic fluctuations with respect to
the GCE. 

The saddle point of the partition function turns out to coincide with the
hydrostatic treatment of the self-gravitating gas \cite{sas,bt} (which is 
usually known as the `isothermal sphere' in the spherically symmetric case).

We find that the {\bf Monte Carlo} simulations (describing thermal equilibrium)
are much more {\bf efficient} than the $N$-body simulations integrating 
Newton's equations of motion. (Indeed, the integration of Newton's equations
provides much more detailed information than the one needed in thermal 
equilibrium investigations).
Actually, a few hundreds of particles are enough to get
quite accurate results in the Monte Carlo simulations (except near the collapse
points). Moreover, the Monte Carlo results turns to be in {\bf excellent}
agreement with the mean field calculations up to very small corrections 
of the order $ 1/N $. 
Our Monte Carlo simulations are performed in a cubic geometry. The
equilibrium configurations obtained in this manner can thus be called
the {\bf `isothermal cube'}.

In summary, the picture we get from 
our calculations using these three  methods show that the self-gravitating gas
behaves as a perfect gas for  $ \eta \to 0, \; 1/\xi\to 0 $. When $
\eta $ and $ 1/\xi $ grow, the gas becomes denser till it suddenly
condenses into a high density object at a critical point GC, C or MC
depending upon the statistical ensemble chosen. 

$ \eta $ is related with the
Jeans' length $ d_J $ of the gas through $ \eta = 3\;  ( L/d_J )^2
$. Hence, when $ \eta $ goes beyond $ \eta_T $, the length of the
system becomes larger than $ d_J / \sqrt{\eta_T /3} $. 
The collapse at
T in the CE is therefore a manifestation of the Jeans' instability. 

\begin{figure}[t] 
\begin{turn}{-90}
\epsfig{file=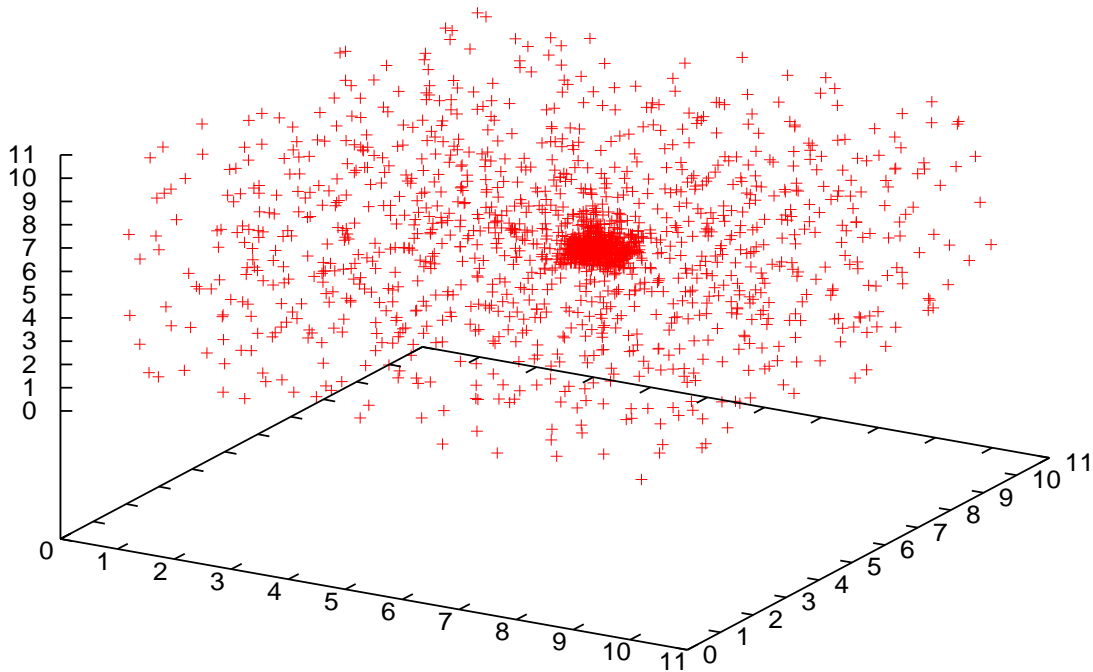,width=12cm,height=18cm} 
\end{turn}
\caption{  Average particle distribution in
the collapsed phase from Monte Carlo simulations with $2000$ particles
in the microcanonical ensemble for $\xi = - 0.6, \; \eta = 0.43, \;
pV/[NT] = 0.414 $.  \label{colmc}}  
\end{figure}

In the MCE, the determinant of fluctuations vanishes at the point MC.
The physical states
beyond MC are collapsed configurations as shown by the Monte Carlo
simulations [see fig. \ref{colmc}]. Actually, the gas collapses in the
Monte Carlo simulations slightly before the mean field prediction for
the point MC. The phase transition at the microcanonical critical point MC is
the so called gravothermal catastrophe \cite{lynbell2}.

The gravitational interaction being attractive without lower
bound, a short distance cut-off ($ A $) must be introduced in order
to give a meaning to the partition function. We take the gravitational
force between particles as $ -G \; m^2 /r^2 $ for $ r > A $ and zero
for $ r< A $ 
where $ r $ is the distance between the two particles. We show that
the cut-off effects are negligible in the $ N = \infty $ limit. That
is, {\bf once} we set $ N = \infty $ with fixed $ \eta $,
{\bf all} physical quantities are {\bf finite} in the
zero cut-off limit ($ A = 0$). The cut-off effects are of the order $
A^2/L^2 $ and can be safely ignored.

In ref. \cite{I} we expressed all global physical quantities in terms of a
single function  $f(\eta)$. Besides
computing numerically  $ f(\eta) $ in the mean field approach,
we showed that this function obeys a
first order non-linear differential equation of first Abel's type.
We obtained analytic results about $ f(\eta) $ from the {\bf Abel's equation}. 
$f(\eta)$ exhibits a {\bf square-root cut} at  $\eta_{C}$, the critical
point in the CE. The first Riemann sheet is realized both
in the CE and the MCE, whereas the second  Riemann sheet (where $ c_V
< 0 $) is only realized in the MCE. $ f(\eta) $ has infinitely
many branches in the $\eta$ plane but only the first two branches are
physically realized. Beyond MC the states described by the mean field
saddle point are unstable. 

We plot and analyze the equation of state, the
energy, the entropy, the free energy, $ c_V $ and the isothermal
compressibility [figs. \ref{fig5}-\ref{fig11}].
Most of these physical magnitudes were not previously
computed in the literature as functions of $ \eta $.

We find analytically the behaviour of $f(\eta)$ near the point $ \eta_C $ in
mean field,
$$
f(\eta) \buildrel{ \eta \uparrow \eta_C}\over= \frac13 +
0.27137\ldots \sqrt{\eta_C-\eta} +0.27763\ldots\; (\eta_C-\eta)
+ O\left[(\eta_C-\eta)^{3/2}\right] \; .
$$
This exhibits a square root {\bf branch point} singularity at $\eta_C$. 
This shows that the specific heat at constant volume diverges  at $\eta_C$ as
$\left(\eta_C-\eta \right)^{-1/2} $ for $ \eta \uparrow
\eta_C$. However,  it must be noticed, that the specific heat at 
constant pressure and the isothermal
compressibility both  diverge at the point $ \eta_T $ as $ \left(\eta_T-\eta
\right)^{-1} $. These mean field results apply for $ |\eta-\eta_C| \ll
1 \ll N|\eta-\eta_C| $. Fluctuations around mean field can be
neglected in such a regime.

The Monte Carlo calculations permit us to obtain  $f(\eta)$ in the
collapsed phase. Such result cannot be
obtained in the mean field approach. The mean field approach only provides
information as $f(\eta)$ in the dilute gas phase. 

For the self-gravitating gas, we find that the Gibbs free energy $ \Phi
$ {\bf is not} equal to $ N $ times the chemical potential and that the
thermodynamic potential $ \Omega $ {\bf is not} equal to $ - PV $ as
usual \cite{llms}. This is a consequence of the dilute thermodynamic
limit $ N \to \infty, \; L\to \infty, \; N/L=$fixed.  

We computed the determinant of small fluctuations 
around the saddle point solution for spherical symmetry in all three
statistical ensembles. 
In the spherically symmetric case, the determinant of small
fluctuations is written as an infinite product over partial waves. The
S and P wave determinants are written in closed form in terms of the
saddle solution. The determinants for higher partial waves are
computed numerically. All partial 
wave determinants are positive definite except for the S-wave \cite{II}. 
The reason why the fluctuations are different in the three ensembles
is rather simple. The more contraints are imposed the smaller becomes
the space of fluctuations. Therefore, in the grand canonical ensemble
(GCE) the system is more free to fluctuate and the phase transition
takes place earlier than in the micro-canonical (MCE) and canonical
ensembles (CE). For the same reason, the transition takes place
earlier in the CE than in the MCE.

The conclusion being that  the mean field correctly gives an excellent
description of the thermodynamic limit except near the critical points
(where the small fluctuations determinant vanishes);
the  mean field is valid for $N|\eta-\eta_{crit}|\gg 1$. The vicinity of the
critical point should be studied in a double scaling limit $N \to
\infty,\; \eta \to \eta_{crit}$. Critical exponents are reported in
ref. \cite{I} for $ \eta \to \eta_C $ using the mean field. These mean field
results apply for $ |\eta-\eta_C| \ll 1 \ll N|\eta-\eta_C| $ with $ N
\gg 1$. Fluctuations around mean field can be neglected in such a regime.

We computed {\bf local} properties of the gas in ref.\cite{II}. That is, the
local energy density $ \epsilon(r) $, local particle density, local
pressure and the local speed of sound. Furthermore, 
we analyze the scaling behaviour of the particle
distribution and its fractal (Haussdorf) dimension \cite{II}.

The particle distribution $ \rho_V({\vec q}) $ proves to be {\bf inhomogeneous}
(except for $ \eta \ll 1 $) and described by an universal function of $ \eta $,
the geometry and the ratio ${\vec r} = {\vec q} / R, \; R$ being the
radial size. Both Monte Carlo simulations and the Mean Field approach
show that the system is inhomogeneous forming a clump of size smaller 
than the box of volume $ V $ [see figs. \ref{gasmc}-\ref{colc}].
 
The particle density in the bulk behaves as 
$ \rho_V({\vec q}) \simeq r^{D-3} $.  That is, the mass $ M(R) 
$ enclosed on a region of size $ R $ vary approximately as
$$
 M(R) \simeq C \;  R^D \; . 
$$
$ D $ slowly decreases from the value $ D = 3 $ for the ideal gas
($\eta=0$) till $ D = 0.98 $ in the extreme limit of the MC point,
$D$ takes the value $1.6$ at $\eta_C$,  [see Table 2]. 
This indicates the presence of a fractal distribution with
Haussdorf dimension $D$. 

Our study of the statistical mechanics of a self-gravitating system indicates 
that gravity provides a dynamical mechanism to produce fractal
structures \cite{I,II,natu}.

The average distance between particles monotonically decrease with $
\eta $ in the first sheet. The mean field and Monte Carlo are very
close in the gaseous phase whereas the Monte Carlo simulations exhibit
a spectacular drop in the average particle distance at the clumping
transition point T. In the second sheet (only
described by the MCE) the average particle distance increases with $
\eta $ \cite{II}. 

We find that the {\bf local} equation of state is given by 
\be \label{pr}
 p(\vec r) = T \, \rho_V(\vec r) \; .
\ee
\noindent We have  {\bf derived} the equation of state for the
self-gravitating gas. It is {\bf locally} the {\bf ideal gas
} equation, but the self-gravitating gas being inhomogeneous,  the
pressure at the surface of a given volume is not equal
to the temperature times the average density of particles in the volume.
In particular, for the whole volume: $ PV/[NT] = f(\eta) \leq 1 $
(the equality holds only  for $ \eta = 0 $).  

Notice that we have found 
the local ideal gas equation of state $ p(\vec r) = T \, \rho_V(\vec r) $ 
for purely gravitational interaction between
particles. Therefore, equations of state different from this one, (as often
assumed and used in the literature for the self-gravitating gas), 
necessarily imply the presence of additional non-gravitational forces. 

The local energy density $\epsilon(r)$ turns out to be an increasing 
function of $r$ in the
spherically symmetric case. The energy density is always positive on
the surface, whereas it is positive at the center for $ 0 \leq \eta  <
\eta_3  = 1.07783\ldots $, and 
negative beyond the point $ \eta = \eta_3  = 1.07783\ldots $.

The local {\bf speed of sound} $ v_s^2(r) $ is computed in the mean 
field approach as a
function of the position for spherical symmetry and long wavelengths. 
$ v_s^2(r) $ diverges at $ \eta  =  \eta_T  = 1.51024\ldots $ in the
first Riemann sheet. Just beyond this point $ v_s^2(r) $ is large and
{\bf negative} in the bulk showing the strongly unstable behaviour of
the gas for such range of values of $ \eta $. 

Moreover, we have
shown the {\bf equivalence} between the statistical mechanical
treatment in the mean field approach and the hydrostatic description
of the self-gravitating gas \cite{sas,bt}.

The success of the hydrodynamical description depends on the value of
the mean free path ($l$) compared with the relevant sizes in the system. 
$ l $ must be $ \ll 1 $. 
We compute   the ratio $ l/a $ ({\bf Knudsen} number), where $ a $ is a
length scale that stays fixed for $ N \to \infty $
and show that  $ l/a \sim N^{-2} $. This
result ensures the accuracy of the hydrodynamical description for
large $ N $. 

Furthermore, we have computed in ref. \cite{II} several
physical magnitudes as functions of $ \eta $ and $ r $ which were not
previously 
computed in the literature as the speed of sound, the energy density,
the average distance between particles and we notice the presence of a
{\bf Haussdorf} dimension in the particle distribution. 

The statistical mechanics of a selfgravitating gas formed by
particles with different masses is thoroughly investigated in
ref.\cite{sieb} while the selfgravitating gas in the presence of 
the cosmological constant is thoroughly investigated
in refs. \cite{negra}. In ref.\cite{cluster} the Mayer expansion
for the selfgravitating gas is investigated in connection with
the stability of the gaseous phase.

The variable $ \eta^R $ appropriate for a spherical
symmetry is defined as 
$$ 
\eta^R \equiv {G \, m^2 N \over R \; T}
= \eta \; \left({4\pi\over 3}\right)^{1/3} =  1.61199\ldots \; \eta \; .
$$               
We use $ \eta^R $ for the mean field theory calculations with spherical
symmetry while $ \eta $ is preferred for the Monte Carlo simulations
with cubic symmetry.

\section{Statistical Mechanics  of the  Self-Gravitating Gas: the
microcanonical and the canonical ensembles}

We investigate a gas of $N$ non-relativistic particles with mass $m$
 self-interacting through Newtonian gravity. We consider first the
 particles isolated, that is, a self-gravitating gas in the
 microcanonical  ensemble. We subsequently consider them in thermal
 equilibrium at temperature $ T \equiv \beta^{-1} $. That is, 
in the canonical ensemble where the
system of $ N $ particles is not isolated but in contact with a thermal bath at
temperature $ T $. 

 We  always assume the system being  on a cubic box of side $ L $ just
 for simplicity. We consider spherical symmetry for the mean field
 approach (sec. 4). Please  notice that we {\bf never} use periodic
 boundary conditions. 

At short distances,  the particle interaction for the self-gravitating
gas in physical situations is not gravitational. Its
exact nature depends on the problem under consideration (opacity limit,
Van der Waals forces for molecules etc.). 
We shall just assume a repulsive short distance potential, that is,
\begin{equation}\label{defva}
v_A(|{\vec q}_l - {\vec q}_j|) = - {1 \over |{\vec q}_l - {\vec q}_j|_A } =
\left\{ \begin{array}{c}  - {1 \over |{\vec q}_l - {\vec q}_j|} \quad
\mbox{for} \; |{\vec q}_l - {\vec q}_j| \ge A \cr 
 \cr +{1 \over A} \quad \mbox{for}\; |{\vec q}_l - {\vec q}_j| \le A
\end{array} \right. 
\end{equation}
where $ A << L $ is the short distance cut-off. 

The presence of the repulsive short-distance interaction prevents the
collapse (here  unphysical) of the self-gravitating gas. In the situations we
are interested to describe (interstellar medium, galaxy distributions)
the collapse situation is unphysical. 

\subsection{The microcanonical ensemble}

The entropy  of the system in can be written as
\begin{equation}\label{smc}
S(E,N)  = \log\left\{ {1 \over N !}\int\ldots \int
\prod_{l=1}^N{{d^3p_l\, d^3q_l}\over{(2\pi)^3}}\; 
\delta\!\left[E - \sum_{l=1}^N\;{{p_l^2}\over{2m}} - 
U({\vec q}_1, \ldots {\vec q}_N) \right] \right\} 
\end{equation}
where
\begin{equation}\label{uq}
U({\vec q}_1, \ldots {\vec q}_N) = - G \, m^2 \sum_{1\leq l < j\leq N} 
{1 \over { |{\vec q}_l - {\vec q}_j|_A}}
\end{equation}
and $G$ is Newton's gravitational constant.

In order to compute the integrals over the momenta $p_l, \; (1 \leq l
\leq N) $, we introduce the variables,
$$
{\vec \rho}_i = {1 \over \sqrt{2m}} \, {\vec p}_i \; .
$$
We can now integrate over the angles in $ 3 N $ dimensions,
\begin{eqnarray}
&&\int_{-\infty}^{+\infty}\ldots \int_{-\infty}^{+\infty}\prod_{l=1}^N\;{
d^3p_l\over{(2\pi)^3}} \;  
\delta\!\left[E - \sum_{l=1}^N\;{\vec\rho}_l^{\, 2} - U({\vec q}_1, \ldots
{\vec q}_N) \right] \cr \cr
&=& \left({\sqrt{2m} \over 2\pi}\right)^{3N}\; {2 \pi^{3N/2} \over
\Gamma\left( {3N \over 2} \right) }\; \int_0^{\infty} \rho^{3N-1}\,
d\rho \; \delta\!\left[E - \rho^2 - U({\vec q}_1, \ldots {\vec q}_N)\right]
\cr \cr
&=& \left( {m \over 2\pi}\right)^{3N/2}{ 1 \over \Gamma\left( {3N
\over 2} \right) } \; \left[E - U({\vec q}_1, \ldots {\vec q}_N)
\right]^{3N/2 -1}  \; \theta\left[E - U({\vec q}_1, \ldots {\vec q}_N) \right] 
\end{eqnarray}
The delta function in the energy thus becomes the constraint of a
positive kinetic energy $ E - U({\vec q}_1, \ldots {\vec q}_N) > 0 $.
We then get for the entropy,
\begin{equation}\label{sq}
S(E,N)  = \log\left\{ {\left({ m \over 2\pi}\right)^{3N/2} \over N !\,
\Gamma\left( {3N \over 2} \right)}\, 
\int_0^L\ldots \int_0^L \prod_{l=1}^N\; d^3q_l \; 
\left[E - U({\vec q}_1, \ldots {\vec q}_N) \right]^{3N/2
-1}\theta\left[E - U({\vec q}_1, \ldots {\vec q}_N ) \right] \right\} 
\end{equation}

It is convenient to introduce the dimensionless variables $ {\vec r}_l
,\;  1\leq l \leq N $ making explicit the volume dependence  as
\begin{eqnarray}\label{variar}
{\vec q}_l &=& L \; {\vec r}_l \quad , \quad {\vec r}_l =(x_l,y_l,z_l)
\;, \cr \cr
0&\leq& x_l,y_l,z_l \leq 1\; .
\end{eqnarray}
That is, in the new coordinates the gas is inside a cube of unit volume.

The entropy then becomes
\begin{eqnarray}\label{sN}
S(E,N)  &=& \log\left\{ {N^{3N-2} \, m^{9N/2-2} \, L^{3N/2 +1} \, G^{3N/2 -1}
\over N !\, \Gamma\left( {3N \over 2} \right)\, {(2\pi)}^{3N/2}}\right.\\ \cr
&&\left.
\int_0^1\ldots \int_0^1 \prod_{l=1}^N\; d^3r_l \; 
\left[\xi + {1 \over N}u({\vec r}_1, \ldots, {\vec r}_N) \right]^{3N/2
-1}\theta\left[\xi + 
{1 \over N}u({\vec r}_1, \ldots, {\vec r}_N) \right] \right\} \nonumber
\end{eqnarray}
where we introduced the  dimensionless variable $\xi$,
\begin{equation}\label{tzi}
\xi \equiv { E \, L \over G \, m^2 \, N^2}
\end{equation}
and
\begin{equation}\label{defu}
u({\vec r}_1, \ldots, {\vec r}_N) \equiv {1 \over
N}\sum_{1\leq l < j\leq N} {1 \over { |{\vec r}_l - {\vec r}_j|_a}}\; .
\end{equation}
where $ a \equiv A/L \ll 1 $. 

Let us define the coordinate partition function in the microcanonical
ensemble as
\begin{equation}\label{defw}
w(\xi,N)\equiv \int_0^1\ldots \int_0^1 \prod_{l=1}^N\; d^3r_l \; 
\left[\xi + {1 \over N}u({\vec r}_1, \ldots, {\vec r}_N) \right]^{3N/2
-1}\theta\left[\xi + 
{1 \over N}u({\vec r}_1, \ldots, {\vec r}_N) \right]\; .
\end{equation}
Therefore,
$$
S(E,N)  = \log\left[ {N^{3N-2} \, m^{9N/2-2} \, L^{3N/2 +1} \, G^{3N/2 -1}
\over N !\, \Gamma\left( {3N \over 2} \right)\, {(2\pi)}^{3N/2}}\right]
+ \log w(\xi,N) \; .
$$
We can now compute the thermodynamic quantities, temperature and
pressure through the standard thermodynamic relations
\begin{equation}\label{termo}
{1 \over T} = \left( {\partial S \over \partial E} \right)_V \quad
\mbox{and} \quad p = T \left( {\partial S \over \partial V} \right)_E
\; ,
\end{equation}
where $ V \equiv L^3 $ stands for the volume of the system and $ p $
is the external pressure on the system.

\subsection{The canonical ensemble}

The partition function in the canonical ensemble can be written as
\begin{equation}\label{fp}
{\cal Z}_C(N,T) = {1 \over N !}\int\ldots \int
\prod_{l=1}^N\;{{d^3p_l\, d^3q_l}\over{(2\pi)^3}}\; e^{- \beta H_N}
\end{equation}
where
\begin{equation}\label{hamic}
H_N = \sum_{l=1}^N\;{{p_l^2}\over{2m}} - G \, m^2 \sum_{1\leq l < j\leq N}
{1 \over { |{\vec q}_l - {\vec q}_j|_A}}
\end{equation}
$G$ is Newton's gravitational constant.

Computing the integrals over the momenta $p_l, \; (1 \leq l \leq N) $

$$
\int_{-\infty}^{+\infty}\;{{d^3p}\over{(2\pi)^3}}\; e^{- {{\beta
p^2}\over{2m}}} = \left({m \over{2\pi \beta}}\right)^{3/2}
$$

yields

\begin{equation}\label{gfpc}
\displaystyle{
{\cal Z}_C(N,T) = {1 \over N !} \left({m \over{2\pi \beta}}\right)^{\frac{3N}2}
\; \int_0^L\ldots \int_0^L
\prod_{l=1}^N d^3q_l\;\; e^{ \beta G \, m^2 \sum_{1\leq l < j\leq N}
{1 \over { |{\vec q}_l - {\vec q}_j|_A}} }}\; .
\end{equation}
We make now explicit the volume dependence introducing the 
variables $ {\vec r}_l ,\;  1\leq l \leq N $ defined in eq.(\ref{variar}).
The partition function takes then the form,
\begin{equation}\label{fp2}
{\cal Z}_C(N,T) = {1 \over N !}\left({m T L^2\over{2\pi}}\right)^{\frac{3N}2}
\; \int_0^1\ldots \int_0^1
\prod_{l=1}^N d^3r_l\;\; e^{ \eta \; u({\vec r}_1,\ldots,{\vec r}_N)}\; ,
\end{equation}
where we introduced the  dimensionless variable $ \eta $ 
\begin{eqnarray}\label{defeta}
\eta &\equiv& {G \, m^2 N \over L \; T}
\end{eqnarray}
and $ u( {\vec r}_1, \ldots,{\vec r}_N) $ is defined by eq.(\ref{defu}).
Recall that
\begin{equation}\label{Upot}
 U \equiv - {G \, m^2 N \over L}\; u({\vec r}_1,\ldots,{\vec r}_N) \; ,
\end{equation}
is the potential energy of the gas.

The free energy  takes then the  form,
\begin{equation}\label{flib}
F = -T \log {\cal Z}_C(N,T) = -N T \log\left[ { e V \over N} \left({mT\over
2\pi}\right)^{3/2}\right]  - T \; \Phi_N(\eta) \; ,
\end{equation}
where
\begin{equation}\label{FiN}
\Phi_N(\eta) = \log \int_0^1\ldots \int_0^1
\prod_{l=1}^N d^3r_l\;\; e^{ \eta \; u({\vec r}_1,\ldots,{\vec r}_N)}\; ,
\end{equation}
The derivative of the function $ \Phi_N(\eta) $ will  be computed by
Monte Carlo simulations, mean field methods and,  in the weak field
limit $ \eta << 1 $, it will be calculated analytically.

We get for the pressure of the gas,
\begin{equation}\label{pres}
p = - \left({ \partial F \over  \partial V}\right)_T = {N T \over V} -
{\eta \, T \over 3 \, V} \; \Phi_N'(\eta)\; .  
\end{equation}
[Here, $ V \equiv L^3 $ stands for the volume of the box and $ p $
is the external pressure on the system.].
We see from eq.(\ref{FiN}) that $ \Phi_N(\eta) $ increases with $ \eta
$ since $ u(.) $ is positive. Therefore, the second term in
eq.(\ref{pres}) is a {\bf negative} correction to the perfect gas
pressure $ {N T \over V} $.  

The mean value of the potential energy $ U $ can be written from
eq.(\ref{Upot}) as
\begin{equation}\label{umed}
<U> = - T \eta  \; \Phi_N'(\eta)
\end{equation}
Combining eqs.(\ref{pres}) and (\ref{umed}) yields the virial theorem,
\begin{equation}\label{virial2}
{p V \over N T} = 1 +{ <U>\over 3 N T}\quad \mbox{or} \quad
{p V \over N T} = \frac12 + { E \over 3 N T} \; ,
\end{equation}
where we use that the average value of the kinetic energy of the gas
is $ \frac32 NT $.

A more explicit form of the equation of state is
\begin{equation} \label{estado}
{p V \over N T} = 1- {\eta \over 3 N} \; \Phi_N'(\eta)\; ,
\end{equation}
where
\begin{eqnarray}\label{fiprima}
\Phi_N'(\eta) &=& e^{-\Phi_N(\eta)} \; \int_0^1\ldots \int_0^1
\prod_{l=1}^N d^3r_l\;  u({\vec r}_1,\ldots,{\vec r}_N)\; e^{ \eta
u({\vec r}_1,\ldots,{\vec r}_N)}\cr 
\cr  &=& \frac12(N-1)  \; e^{-\Phi_N(\eta)} \; \int_0^1\ldots \int_0^1
\prod_{l=1}^N d^3r_l\;{1 \over |{\vec r}_1-{\vec r}_2|_a}\; e^{ \eta
u({\vec r}_1,\ldots,{\vec r}_N)} \; .
\end{eqnarray}
This formula indicates that $ \Phi_N'(\eta) $ is of order $ N $ for
large $ N $. Monte Carlo simulations  as well as analytic calculations
for small $ \eta $  show that this is indeed the case. In conclusion,
we can write the equation of state of the self-gravitating gas as
\begin{equation}\label{pVnT}
{p V \over N T} = f(\eta) \quad ,     
\end{equation}
where the function 
$$ 
f(\eta) \equiv  1- {\eta \over 3 N} \; \Phi_N'(\eta)\; ,
$$ 
is {\bf independent} of $ N $  for large $N$ and fixed $ \eta $.
[In practice, Monte Carlo simulations show that $ f(\eta) $ is
independent of $ N $ for  $ N > 100 $]. 

We get in addition,
\begin{equation}\label{Ueta}
<U>= -3 N T\;[ 1-  f(\eta)]\;.
\end{equation}
In the dilute  limit, $ \eta \to 0 $ and we find the perfect gas
value
$$
f(0) = 1\; .
$$
Equating eqs.(\ref{estado}) and (\ref{pVnT}) yields,
$$
\Phi_N(\eta)= 3N \, \int_0^{\eta} dx \, { 1 - f(x) \over x}\; .
$$
Relevant thermodynamic quantities can be expressed in terms of the
function $ f(\eta) $. We find for the free energy from
eq.(\ref{flib}),
\begin{equation}\label{enlib}
F =  F_0- 3NT\; \int_0^{\eta} dx \; { 1 - f(x) \over x}\; . 
\end{equation}
where 
\begin{equation}\label{Fcero}
F_0 = -N T \log\left[{ e V \over N} \left({mT\over
2\pi}\right)^{3/2}\right]
\end{equation}
is the free energy for an ideal gas.

We find for the total energy $ E $, chemical potential $ \mu $ and
entropy $ S $ the following expressions,
\begin{eqnarray}\label{ecan}
E &=&  3NT\left[  f(\eta) -\frac12\right] \; , \\  \cr
\label{poqui}
\mu &=& \left({ \partial F \over  \partial N}\right)_{T,V} = 
-T \log\left[{V \over N}\left({mT\over 2\pi}\right)^{3/2}\right]
- 3T[1 - f(\eta)] - 3T\; \int_0^{\eta} dx \, { 1 - f(x) \over x} 
\; , \cr \cr
\label{entro}
S &=& - \left({ \partial F \over  \partial T}\right)_V \cr \cr
&=& S_0  + 3N\left[
\int_0^{\eta} dx \, { 1 - f(x) \over x}+  f(\eta) -1 \right] \; ,
\end{eqnarray}
where
$$
S_0 = -{F_0 \over T} +\frac32 N \; . 
$$
is the entropy of the ideal gas.

Notice that here the Gibbs free energy 
\begin{equation}\label{gibbs}
\Phi = F + pV = F_0 + NT \left[ f(\eta) - 3 \,  \int_0^{\eta} dx \, {
1 - f(x) \over x}  \right] \; ,
\end{equation}
is {\bf not} proportional to the chemical potential. That is,
here $ \Phi \neq \mu \, N $ and we have instead,
\begin{equation}\label{gibbs2}
\Phi - \mu \, N = 2 NT \left[ 1-f(\eta) \right] \; .
\end{equation}
This relationship differs from the customary one (see \cite{llms}) due
to the fact that the dilute scaling relation $ N \sim L $ holds here
instead of the usual one $ N \sim L^3 $. The usual relationship is
only recovered in the ideal gas limit $ \eta = 0 $.  

\bigskip

The specific heat at constant volume takes the form\cite{llms},
\begin{eqnarray}\label{ceV}
c_V &=& {T \over N}  \left({ \partial S \over  \partial T}\right)_V
= 3 \left[  f(\eta)-\eta \; f'(\eta) -\frac12 \right]\; .
\end{eqnarray}
where we used eq.(\ref{entro}).
This quantity is also related to the fluctuations of the potential
energy $(\Delta U)^2$ and it is positive defined in the canonical ensemble,
\begin{equation}\label{cV}
c_V =\frac32 + (\Delta U)^2 \; .
\end{equation}
Here,
\begin{equation}\label{delu}
(\Delta U)^2 \equiv {{<U^2>-<U>^2}\over N \; T^2}= 3 \left[
f(\eta)-\eta \; f'(\eta) - 1 \right]\; .
\end{equation}

The specific heat at constant pressure is given by \cite{llms}
\begin{equation}\label{cpcv}
c_P = c_V -  {T \over N}{{ \left({ \partial p \over  \partial
T}\right)^2_V}\over { \left({ \partial p \over  \partial V}\right)_T}}\; .
\end{equation}

and then,
\begin{eqnarray}\label{ceP}
c_P &=& c_V + { \left[f(\eta)-\eta f'(\eta)\right]^2 \over
f(\eta)+\frac13 \eta f'(\eta)} \cr \cr
&=& -\frac32 + {{4\, f(\eta)\left[ f(\eta)-\eta f'(\eta)\right] } \over
{f(\eta)+\frac13 \eta f'(\eta)}} \; .
\end{eqnarray}

The isothermal ($K_T$) and adiabatic ($K_S$) compressibilities take the form
\begin{eqnarray}\label{KT}
K_T &=& - { 1 \over V} \left({ \partial V \over  \partial p}\right)_T =
{V \over N\, T} {1 \over {f(\eta)+\frac13 \eta f'(\eta)}} \; ,\cr \cr
K_S &=& - { 1 \over V} \left({ \partial V \over  \partial p}\right)_S = 
{ c_V \over c_P} \; K_T \; .
\end{eqnarray}
It is then convenient to introduce the compressibilities
\begin{eqnarray}\label{kapa}
\kappa_T \equiv {NT \over V} \, K_T = {1 \over {f(\eta)+\frac13 \eta
f'(\eta)}} \quad \mbox{and} \quad \kappa_S \equiv{NT \over V} \, K_S = 
{ c_V \over c_P} \; \kappa_T \; ,
\end{eqnarray}
which are both of order one (intensive) in the $ N, \; L \to \infty $
limit with $ N/L $ fixed. 

\bigskip

The speed of sound $ v_s $ can be written as \cite{llmf}
\begin{equation}\label{defson}
v_s^2 = - {{ c_P \; V^2} \over {c_V \; N}} \left({ \partial p \over
\partial V}\right)_T = {V^2 \over N} \left[ {T \over N \, c_V}  \left({
\partial p \over  \partial T}\right)^2_V  - \left({ \partial p \over
\partial V}\right)_T \right] \; .
\end{equation}
where we used eq.(\ref{cpcv}) in the last step. 
Therefore,
\begin{equation}\label{vson}
{v_s^2 \over T} = { \left[f(\eta)-\eta f'(\eta)\right]^2\over 3
\left[f(\eta)-\eta f'(\eta)-\frac12 \right]} + f(\eta)+\frac13 \eta
f'(\eta)\; .
\end{equation}
The pressure $ p $ used in this calculation corresponds to the
pressure on the surface of the system. Hence, this is the speed of
sound on the surface of the system, this is different from the
speed of sound 
inside the volume since the ground state is inhomogeneous. We compute
the speed of sound as a function of the point in \cite{II}.

\bigskip

We see that the large $ N $ limit of the self-gravitating gas is
special. Energy, free energy  and entropy are {\bf extensive} magnitudes
in the sense that they are proportional to the number of particles $ N
$ (for fixed $\eta$). 
They all depend on the variable $ \eta ={G \, m^2 N \over L \; T} $
which is to be kept fixed for the thermodynamic limit
($ N\to \infty $ and $ V \to \infty $) to exist.
Notice that  $ \eta $ contains the
ratio $ N/L = N \; V^{-1/3} $ which must be considered here an 
{\bf intensive variable}. Here, the presence of long-range gravitational
situations calls for this new intensive variable in the thermodynamic limit.

In addition, all physical magnitudes can be expressed in terms of a
single function of one variable: $ f(\eta) $.

\section{Monte Carlo Simulations}

The Metropolis algorithm\cite{montec} was applied for the first time to 
the self-gravitating gas in ref.\cite{I}. 
We computed in this way the pressure, the energy, the average density,
the potential energy fluctuations, the average particle distance
and the average squared particle distance as functions of $ \eta $
for the self-gravitating gas in a cube of size $ L $ in the canonical
ensemble at temperature $ T $\cite{I}.
 
We implemented the Metropolis algorithm in the following way. We start from a 
random distribution of $ N $ particles in the chosen volume. We update such
configuration choosing a particle at random and changing at random its 
position. We then compare the energies of the former and the new 
configurations. We use the standard Metropolis test to choose between the new 
and the former configurations. The energy of the configurations
are calculated performing the exact sums as in eq.(\ref{defu}). We use as 
statistical weight for the Metropolis algorithm in the canonical ensemble,
$$
 e^{ \eta \; u({\vec r}_1,\ldots,{\vec r}_N)}\; ,
$$
which appears in the coordinate partition function eq.(\ref{FiN}).
The number of particles $ N $ went  up to $ 2000 $. 

We introduced
a small short distance cutoff $ A = 10^{-4}L  - 10^{-8} L$ in the
attractive Newton's potential according to eq.(\ref{defva}). All
results in the gaseous phase were insensitive to the cutoff value. 
The partition function calculation turns to be much less sensible to
the short distance singularities of the gravitational force than
Newton's equations of motion for $N$ particles. That is, solving the
classical dynamics for $N$ particles interacting through gravitational
forces as well as solving the Boltzman equation including the $N$-body
gravitational interaction requires sophisticated algorithms to avoid
excessively long computer times. As is clear, solving the $N$-body
classical evolution or the kinetic equations provides the
time-dependent dynamics and out of thermal equilibrium effects which
are out of the scope of our approach.  

\begin{figure}[t] 
\begin{turn}{-90}
\epsfig{file=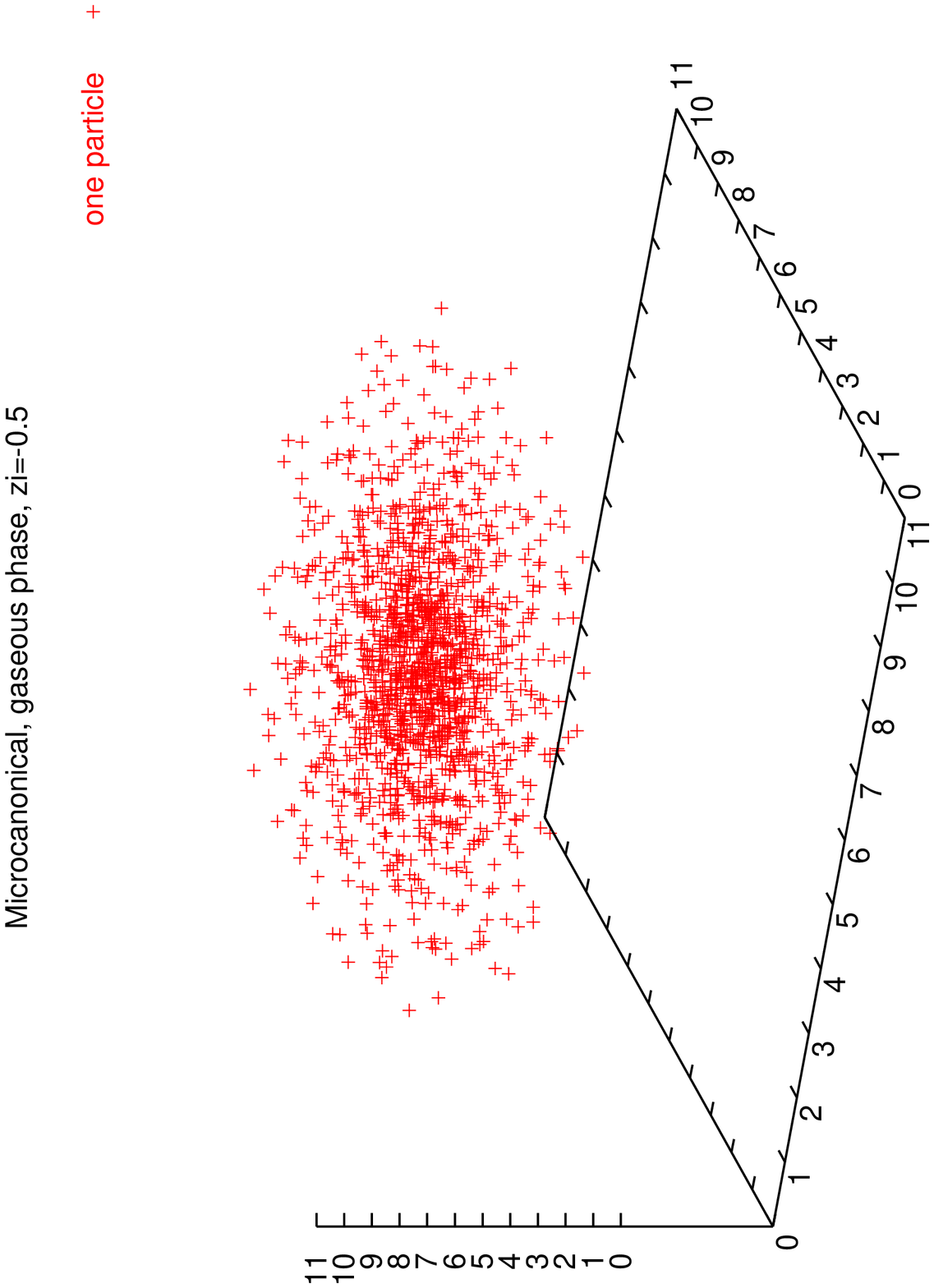,width=12cm,height=18cm} 
\end{turn}
\caption{ Average particle distribution in
the gaseous phase from Monte Carlo simulations with $2000$ particles
in the microcanonical ensemble for $\xi = - 0.5, \; \eta = 1.38, \;
pV/[NT] = 0.277$.  \label{gasmc}}  
\end{figure}

In the CE, two different phases show up: for $ \eta < \eta_T $ we have a
non-perfect gas and for $  \eta > \eta_T $ it is  a condensed
system with {\bf negative} pressure. The transition between  the two phases
is very sharp. This phase transition is associated with the Jeans
instability.

A negative pressure indicates that the free energy grows for
increasing volume at constant temperature [see
eq.(\ref{pres})]. Therefore, the system wants to contract sucking on
the walls.   

We plot in fig. \ref{fig14} $ f(\eta) = pV/[NT] $ as function of $ \eta $.  
$ pV/[NT] $ monotonically decreases with  $ \eta $.

We find an excellent agreement between the the Monte Carlo  and the
mean field results for $ pV/[NT] $ in the gaseous phase plotted in fig.
\ref{fig5}. This excellent agreement appears for $N$ as low as $500$.

In the Monte Carlo simulations the phase transition to the condensed
phase happens for $ \eta = \eta _T $ slightly below $ \eta_C $.
For $ N = 2000 $ we find $ \eta_T \sim 1.515 $.  
For $ \eta _T < \eta < \eta_C $, the gaseous phase only exists as a
 short lived metastable state as seen in our simulations.

The average distance between particles $ <r> $  and the average
squared distance between particles $ <r^2> $ monotonically decrease
with  $ \eta $. When the gas collapses at   $ \eta_T \; , <r> $
and $ <r^2> $ exhibit a sharp decrease.  

The values of $ pV/[NT] , \;  <r> $ and $ <r^2> $ in the condensed phase are
independent of the cutoff for $ a < 10^{-5} $. The Monte Carlo results
in this condensed phase can be approximated for 
$ \eta > 2 $ 
as 
\begin{equation}\label{pcolap}
{pV \over NT} = f(\eta) \simeq  1 -  K \; \eta \quad ,
\quad  <r> \simeq 0.016  \; .
\end{equation}
where $ K \simeq 14 $. 

\bigskip

Since $ f(\eta) $ has a jump at the transition, the Gibbs free energy
$ \Phi $ is discontinuous and we have a phase
transition of the {\bf zeroth} order. We find from eq.(\ref{gibbs})
\begin{equation}\label{deltaG}
{\Phi(\mbox{collapse})-\Phi(\eta_T)  \over N \, T} =
f(\mbox{collapse}) - f(\eta_T) \simeq -21  < 0 \; . 
\end{equation}

We can easily compute the latent heat of the transition per particle ($q$)
using the fact that the volume $V$ stays constant. Hence, $ q = \Delta
E/N $ and we obtain from eq.(\ref{ecan})
\begin{equation}\label{qsobreT}
{q \over T} = {E(\mbox{collapse}) -E(\eta_T)  \over N \, T} =3 \; \left[
f(\mbox{collapse}) - f(\eta_T) \right] 
\simeq 2 - 3  \, K \; \eta_T\simeq -62  < 0 \; . 
\end{equation}
This phase transition is different from the usual phase transitions
since the two phases cannot coexist in equilibrium as their pressures
are different.

Eq.(\ref{pcolap}) can be understood from the general treatment in
sec. III as follows. We have from eqs.(\ref{estado})-(\ref{fiprima}) 
\begin{equation}\label{Feta}
f(\eta) = 1 - {\eta \over 3} < {1 \over r}> \; .
\end{equation}
The Monte Carlo results indicate that $  < {1 \over r}> \simeq 42 $ is
approximately constant in the collapsed region as well as $ <r> $ and
$ <r^2> $.  Eq.(\ref{pcolap}) thus follows from eq.(\ref{Feta}) using
such value of $ < {1 \over r}> $. 

\bigskip

The behaviour of $ pV/[NT] $ near  $ \eta_C $ in the gaseous phase 
can be well reproduced by 
\begin{equation}\label{pecrit}
{pV \over NT}= f(\eta) \;\; {\buildrel{ \eta \uparrow \eta_C}\over =}\;
 f_C + A \; \sqrt{\eta_C -\eta} 
\end{equation}
where $  f_C \simeq 0.316, \; A \simeq 0.414 $ and  $ \eta_C \simeq 1.540 $.

In addition, the
behaviour of $ (\Delta U)^2 $ in the same region is well reproduced by
\begin{equation}\label{flUc}
(\Delta U)^2\; \; {\buildrel{ \eta \uparrow \eta_C}\over =}\;\; C  + 
{D \over \sqrt{\eta_C-\eta} }
\end{equation}
with $ C \simeq -1.64 $  and $ D \simeq 0.901 $. 
[Notice that for finite $ N , \; (\Delta U)^2 $ will be finite albeit
very large at the phase transition]. Eq.(\ref{delu}) relating $
f(\eta) $ and $ (\Delta U)^2 $ is satisfied with reasonable
approximation. 

We thus find a critical region just below $ \eta_C $ where the energy
fluctuations tend to infinity as $ \eta \uparrow \eta_C $.

The point $ \eta_T $ where the phase transition actually takes place
in the Monte Carlo simulations is at $ \eta_T \simeq 1.51 < \eta_C
$. This value for $ \eta_T $ is numerically close to the point 
where the isothermal compressibility $ \kappa_T $ diverges
and becomes negative. As stressed in ref.\cite{I,II} these two points
actually coincide. The collapse observed in the Monte Carlo
simulations corresponds to the singularity of  $ \kappa_T $ where, in
addition,  the speed of sound at the origin becomes imaginary\cite{I,II}.

Since Monte Carlo simulations are like real experiments, we conclude
that the gaseous phase extends from $ \eta = 0 $ till $ \eta = \eta_T
$  in the CE and {\bf not} till $ \eta = \eta_C $. 
Notice that in the literature based on the hydrostatic
description of the self-gravitating gas \cite{sas,pad,HK,bt}, only the
instability at $ \eta = \eta_C $ is discussed whereas the
singularities at $ \eta = \eta_T $ are not considered.

\bigskip

We then performed Monte Carlo calculations in the microcanonical
ensemble where the coordinate partition function is given by eq.(\ref{defw}). 
We thus used  
$$
\left[\xi + {1 \over N}\, u({\vec r}_1, \ldots, {\vec r}_N) \right]^{3N/2
-1}\theta\left[\xi + 
{1 \over N} \, u({\vec r}_1, \ldots, {\vec r}_N) \right]\; ,
$$
as the statistical weight for the Metropolis algorithm. 

The MCE and CE Monte Carlo results coincide up to the
statistical error for $0 < \eta <\eta_T $, that is for $ \infty >
\xi > \xi_T \simeq -0.19 $. In the MCE the gas does not clump at $\eta
= \eta_C $ (point $C$ in fig. \ref{fig14}) and the specific heat
becomes negative 
between the points $C$ and $MC$. In the MCE the gas does clump at
$ \xi \simeq -0.52 \; , \; \eta^T_{MC} \simeq 1.33 $ (point $MC$ in
fig. \ref{fig14}) increasing  {\bf both its temperature and pressure
discontinuously}. We find from the Monte Carlo data that the
temperature increases by a factor $ 2.4 $ whereas the pressure
increases by a factor $ 3.6 $ when the gas clumps. The transition
point $ \eta^T_{MC} $ in 
the Monte Carlo simulations is slightly to the right of the critical
point $ \eta_{MC} $  predicted by mean field theory. The mean field
yields for the sphere $ \eta_{MC} = 1.2598\ldots $.

\bigskip

As is clear, the domain between $C$ and $MC$ cannot
be reached in the CE since $ c_V > 0 $ in the CE as shown by eq.(\ref{cV}).  

\begin{figure}[t] 
\begin{turn}{-90}
\epsfig{file=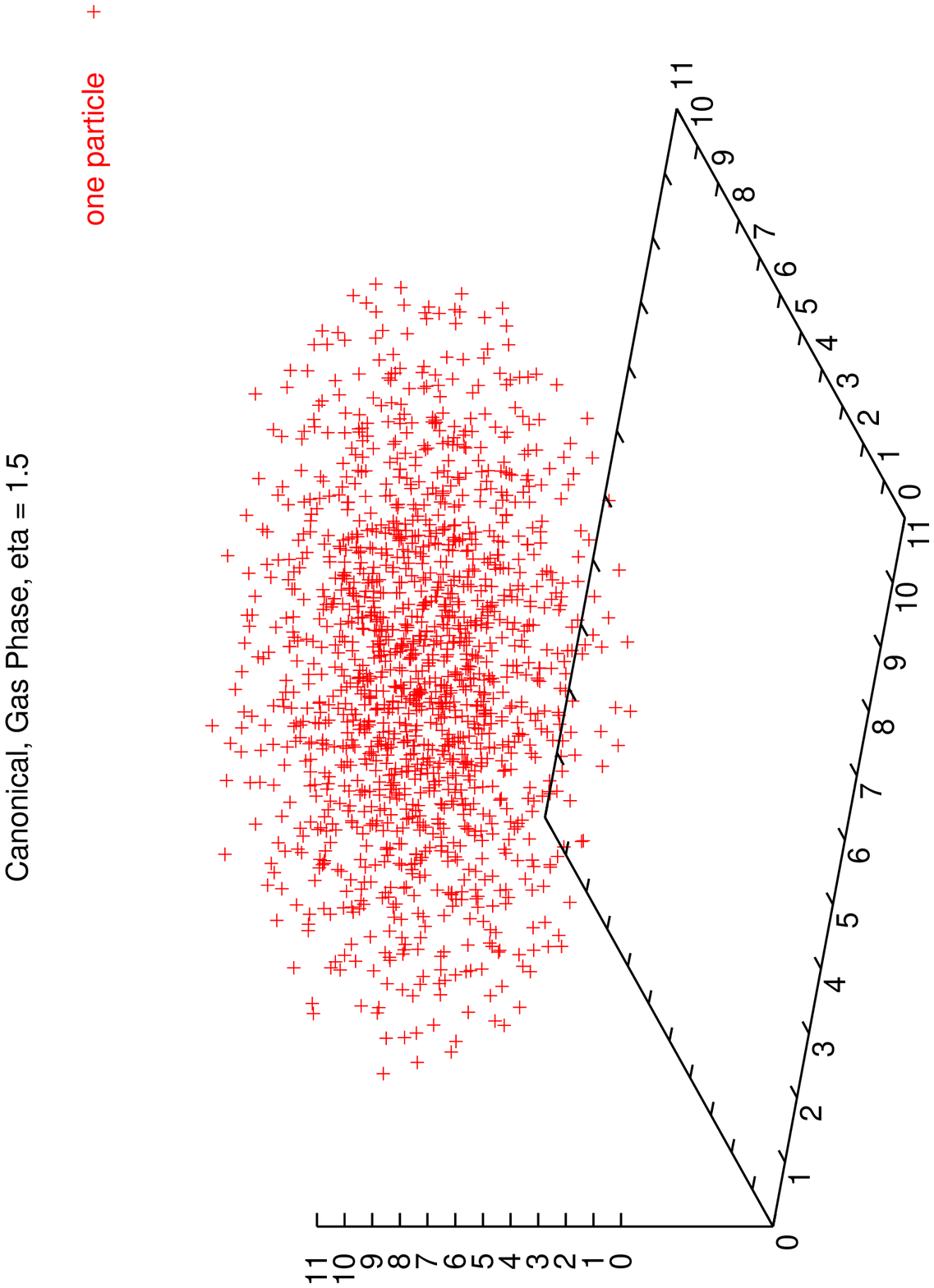,width=12cm,height=18cm} 
\end{turn}
\caption{Average particle distribution in the gaseous phase from Monte
Carlo simulations in the 
canonical ensemble for $ \eta = 1.5 $ and $ N = 2000 $ 
\label{gasc}} 
\end{figure}

We find an excellent agreement between the Monte Carlo  and Mean
Field (MF) results
(both in the MCE and CE). (This happens although the  geometry for the MC
calculation is cubic while it is spherical for the MF). The points
where the collapse phase transition  occurs ($\eta_T$ and $\eta^T_{MC}$) 
slowly increase with the number of particles $N$.

We verified that the Monte Carlo results in the gaseous phase  
($ \eta < \eta_T $) are cutoff independent for $ 10^{-3} \geq a \geq 10^{-7}
$. 

\bigskip

As for the CE, the Gibbs free energy is discontinuous at the
transition in the MCE. The transition is then of the zeroth order.
We find from eq.(\ref{gibbs})
$$
{\Phi(\mbox{collapse})-\Phi(\eta_T)  \over N \, T_{gas}} =
 {T_{coll} \over T_{gas}}\, f(\mbox{collapse}) - f(\eta_T) \simeq
 0.7  > 0 \; .  
$$
where we used the numerical values from the Monte Carlo simulations.
Notice that the Gibbs free energy {\bf increases} at the MC transition
whereas it decreases at the C transition [see eq.(\ref{deltaG})]. 

Here again
the two phases cannot coexist in equilibrium since their pressures and
temperatures are different.

\bigskip

We display in figs. \ref{gasmc}-\ref{colmc} the average particle
distribution from 
Monte Carlo simulations with $2000$ particles in the microcanonical
ensemble at both sides of the gravothermal catastrophe, i. e. $ \eta =
\eta_{MC} $. Fig. \ref{gasmc} corresponds to the gaseous phase and
fig. \ref{colmc} to the collapsed phase. The inhomogeneous particle
distribution is clear in fig. \ref{gasmc} whereas fig. \ref{colmc}
shows a dense collapsed core surrounded by a halo of particles. 

The different nature of the collapse in the CE and in the MCE can be
explained using the virial theorem [see eq.(\ref{virial2})]
$$
{p \, V \over N \, T} = 1 + { U \over N \, T} \; .
$$
When the gas collapses in the CE the particles get very close and $ U
$ becomes large and negative while $ T $ is fixed. Therefore, $ p \, V
\over N \, T $ may become large and negative as it does.

We can write the virial theorem also as,
$$
p \, V - \frac12 \, NT = \frac13 \, E\; .
$$
When the gas is near the point MC, $ E < 0 $ is fixed and we have $ T
> 0 $. Therefore, $ p \, V \over N \, T $ as well as $ U = E - 3\, N \,
 T / 2 $ cannot become large and negative as in the CE collapse.
This prevents the distance between the particles to
decrease. Actually, the Monte Carlo simulations show that $ <r> $ {\bf
increases} by $ 18\% $ when the gas collapses in the MCE. 

\begin{figure}[t] 
\begin{turn}{-90}
\epsfig{file=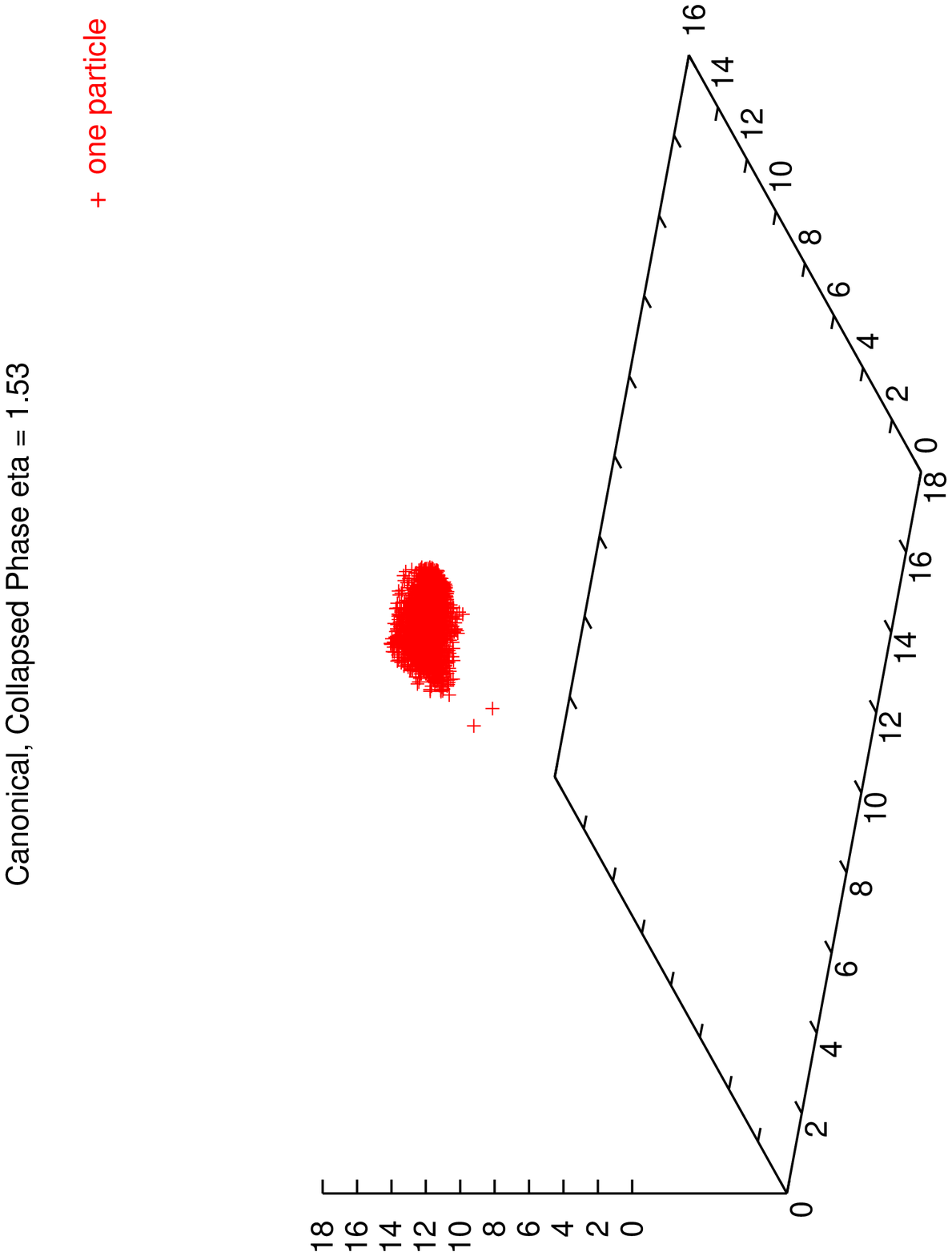,width=12cm,height=18cm} 
\end{turn}
\caption{  Average particle distribution in
the collapsed phase from Monte Carlo simulations with $2000$ particles
in the canonical ensemble for $ \eta = 1.53, \;
pV/[NT] = -14.44 $.  \label{colc}}  
\end{figure}

Figs. \ref{gasc} and \ref{colc} depict the average particle
distribution from  Monte Carlo simulations with $2000$ particles in
the canonical ensemble at both sides of the collapse critical point,
i. e. $ \eta = 
\eta_{C} $. Fig. \ref{gasc} corresponds to the gaseous phase and
fig. \ref{colc} to the collapsed phase. The inhomogeneous particle
distribution is clear in fig. \ref{gasc} whereas fig. \ref{colc}
shows a dense collapsed core surrounded by a very little halo of particles.

Notice that the collapsed phases are of different nature in the CE and
MCE. The core is much tighter and the halo much smaller  in the CE
than in the MCE.

Figs. 3 and 5 depict the average particle distribution for the gaseous
phase in the MCE and the CE, respectively. In this phase, the MC
simulations  give identical descriptions for large $N$ in both
ensembles. [This important point will be further demostrated in
sec. VI by functional integral methods]. The average configurations in
figs. 3 and 5 describe a self-gravitating gas in thermal equilibrium
within a {\bf cube}. We may call it the {\bf isothermal cube} by
analogy with the well known isothermal sphere\cite{emden}-\cite{bt}.

\section{Mean Field Approach}

Both in the microcanonical and the canonical ensembles the coordinate
partition functions are given by $3N$-uple integrals [eqs.(\ref{defw})
and (\ref{FiN}), respectively]. In the $ N \to \infty $ limit both $3N$-uple
integrals can be recasted as functional integrals over the continuous
particle density as shown in refs.\cite{I,II}. We present here the
mean field approach in the canonical ensemble. The mean field approach
for the microcanonical and grand canonical ensembles can be find in
refs.\cite{I,II}. 

The coordinate partition function $ e^{\Phi_N(\eta)} $ 
given by eq.(\ref{FiN}) can be recasted as a functional integral in
the thermodynamic limit.  
\begin{eqnarray}\label{zcanmf}
e^{\Phi_N(\eta)} &\buildrel{ N>>1}\over=& \int\int D\rho\; d{\hat a} \;
e^{-N s_C[\rho(.),\hat a,\eta]} \\ \cr
s_C[\rho(.),\hat a,\eta] &=& -{\eta \over2} \int  {d^3x  \; d^3y \over |{\vec
x}-{\vec y}|} \;\rho({\vec x}) \; \rho({\vec y}) + \int d^3x \; \rho({\vec
x}) \; \log\rho({\vec x}) - i{\hat a}\left(\int  d^3x \,
\rho({\vec x}) - 1 \right) \;.\nonumber 
\end{eqnarray}
where we used the coordinates $ \vec x $ in the unit volume.
The first term  is the potential energy, the second term is the functional
integration measure for this case (see appendix A). Here $ N \;
\rho({\vec x}) $ stands for the density of particles. 

The integration over $ \hat a $ enforces the number of particles to be
exactly $ N $:
\begin{equation}\label{vincu}
\int  d^3x \; \rho({\vec x}) = 1
\end{equation}
That is, in the coordinates $ \vec q $ (running from $ 0 $ to $ L $),
the density of particles is
$$
{N \over L^3} \; \rho({\vec q}) \quad {\rm with} \quad\int  d^3q \;{N
\over L^3} \; \rho({\vec q}) = N \; .
$$
The functional integral in  eq.(\ref{zcanmf}) is dominated for large $N$ by
the extrema of the `effective action' $ s_C[\rho(.),\hat a,\eta] $,
that is, the solutions of the stationary point equation
\begin{equation}\label{eqrho}
\log \rho_s({\vec x}) - \eta \int{ d^3y \; \rho_s({\vec y})\over |{\vec
x}-{\vec y}|} =a_s \; ,
\end{equation}
$ a = i{\hat a} $ is a Lagrange multiplier enforcing 
the constraint (\ref{vincu}).

Applying the Laplacian and setting $\phi({\vec x}) \equiv
\log\rho_s({\vec x})$ yields, 
\begin{equation}\label{puntoest}
\nabla^2\phi({\vec x}) +   4\pi \eta \; e^{\phi({\vec x})} = 0 \; ,
\end{equation}
This equation is scale covariant \cite{prd}. That is, if $ \phi({\vec
x}) $ is a solution of eq.(\ref{puntoest}), then
\begin{equation}\label{coves}
\phi_{\lambda}({\vec x}) \equiv \phi(\lambda{\vec x}) +\log\lambda^2
\end{equation}
where $ \lambda $ is an arbitrary constant
is also a solution of eq.(\ref{puntoest}). For spherically symmetric
solutions this property  can be found in ref.\cite{chandra}. 

Integrating eq.(\ref{puntoest}) over the unit volume and using the
constraint (\ref{vincu}) yields
\begin{equation}\label{vincu2}
\int {\vec \nabla} \phi({\vec x}) \cdot d{\vec s} = - 4\pi \eta
\end{equation}
where the surface integral is over the boundary of the unit volume.

In the mean field approximation we only keep the dominant order for
large $ N $. Therefore, only the exponent at the saddle point accounts
and  according to eq.(\ref{flib}) we find for the free energy 
\begin{eqnarray}\label{fcampmed}
F &=& F_0 +  N\, T\, s(\eta) + {\cal O}(N^0)\cr \cr
{ p V \over NT} &=& 1 + {\eta \over 3 } { ds \over d \eta} + {\cal O}(N^{-1})
\end{eqnarray}
Hence, in the mean field approximation, the function $ f(\eta) $ is given by
\begin{equation}\label{fetacm}
f_{MF}(\eta)\equiv 1 + {\eta \over 3 } { ds \over d \eta}\; ,
\end{equation}
From eq.(\ref{zcanmf}) we can compute $ s(\eta) $ in terms of the
saddle point solution as follows
\begin{equation}
s(\eta) \equiv s_C[\rho_s(.),a_s,\eta] = -{\eta \over 2} \int  {d^3x  \; d^3y
\over |{\vec x}-{\vec y}|} \;\rho_s({\vec x}) \; \rho_s({\vec y}) + \int
d^3x \; \rho_s({\vec x}) \; \log\rho_s({\vec x}) \; . 
\end{equation}
Using eq.(\ref{eqrho}) we find the equivalent expression,
\begin{equation}\label{sdeta}
s(\eta)= {a_s \over 2} +
\frac12 \int  \phi({\vec x}) \; e^{ \phi({\vec x})} \; d^3x \; .
\end{equation}

\section{Mean Field Results}

Let us summarize here the main results of ref. \cite{I} in the mean field
approach for spherical symmetry.

The saddle point is given by
\begin{equation}\label{fixi}
\phi(r) = \log \rho(r) = \log\left({\lambda^2 \over  4\, \pi \,\eta^R}\right) +
\chi(\lambda \, r) \; .
\end{equation}
Here $ \rho(r) $ is the particle density and $ \chi(\lambda ) $ obeys
the equation
\begin{equation}\label{ecuaxi}
\chi''(\lambda) + {2 \over \lambda} \, \chi'(\lambda) +
e^{\chi(\lambda)} = 0 \quad , \quad \chi'(0) = 0 \quad , \quad
\chi(0) = 0\ ; .
\end{equation}
$ \chi(x) $ is independent of $ \eta^R $, and  $ \lambda $ is related to
 $\eta^R$ through 
\begin{equation}\label{lambaxi}
\lambda \; \chi'(\lambda ) = -\eta^R \; .
\end{equation}
We have in addition,
\begin{equation}\label{fide1}
\phi(1) = \log\left[ {3 \, f_{MF}(\eta^R) \over 4 \, \pi }\right]
\quad , \quad \rho(1) = {3 \over 4 \, \pi } \; f_{MF}(\eta^R) \; .
\end{equation}

\bigskip
\begin{figure}
\begin{turn}{-90}
\epsfig{file=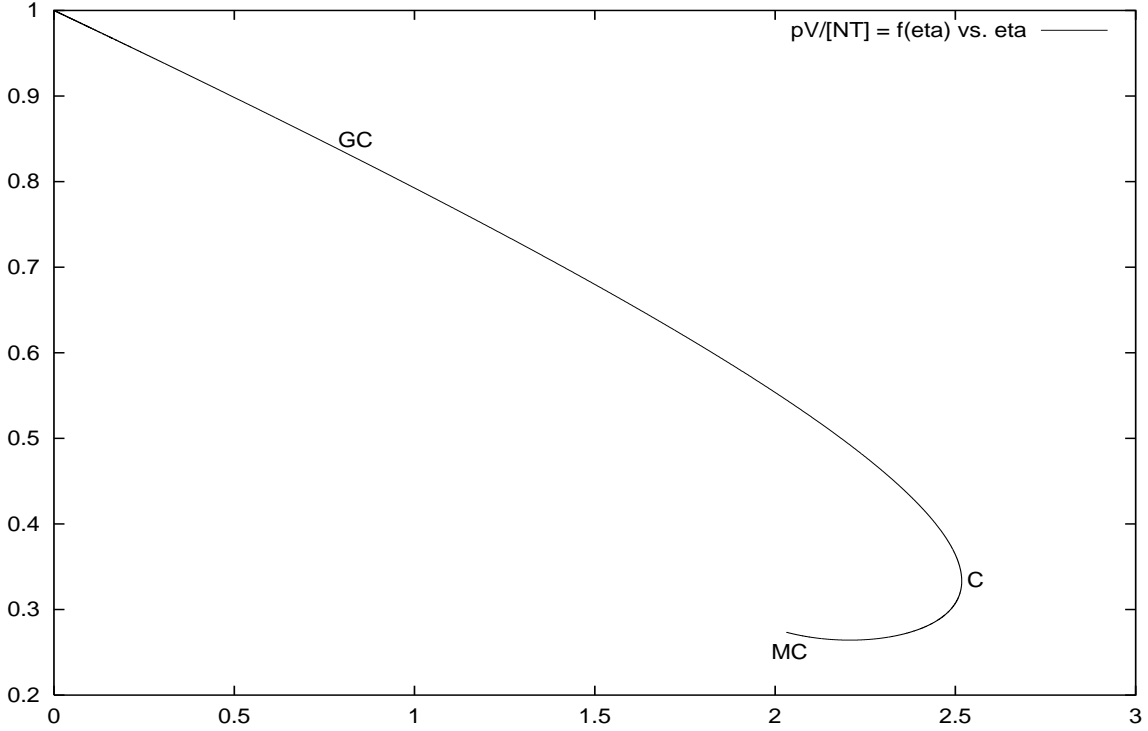,width=10cm,height=16cm} 
\end{turn}
\caption{ $ f_{MF}(\eta^R) = P V/[ N T] $ as a function of $\eta^R$ in the MF
approximation [eq.(\ref{abel})].  $ f_{MF}(\eta^R) $ has a square root
branch point at $ \eta^R_C $. The points GC, C and MC indicate the
transition to the collapsed phase for each ensemble (grand canonical,
canonical and microcanonical, respectively): $ \eta^R_{GC} =
0.797375\ldots, \; \eta^R_{C} = 2.517551\ldots, \; \eta^R_{MC} =
2.03085\ldots $ (notice that $ \eta^R_{MC} $ is in the second Riemann sheet). 
Since $E/[3NT] = f_{MF}(\eta^R) -\frac12 $, this plot also shows
the energy per particle as a function of $\eta^R$. Furthermore, the
particle density at the surface is proportional to $f_{MF}(\eta^R)$
[see eq.(\ref{fide1})].
\label{fig5}} 
\end{figure} 
\begin{figure}
\begin{turn}{-90}
\epsfig{file=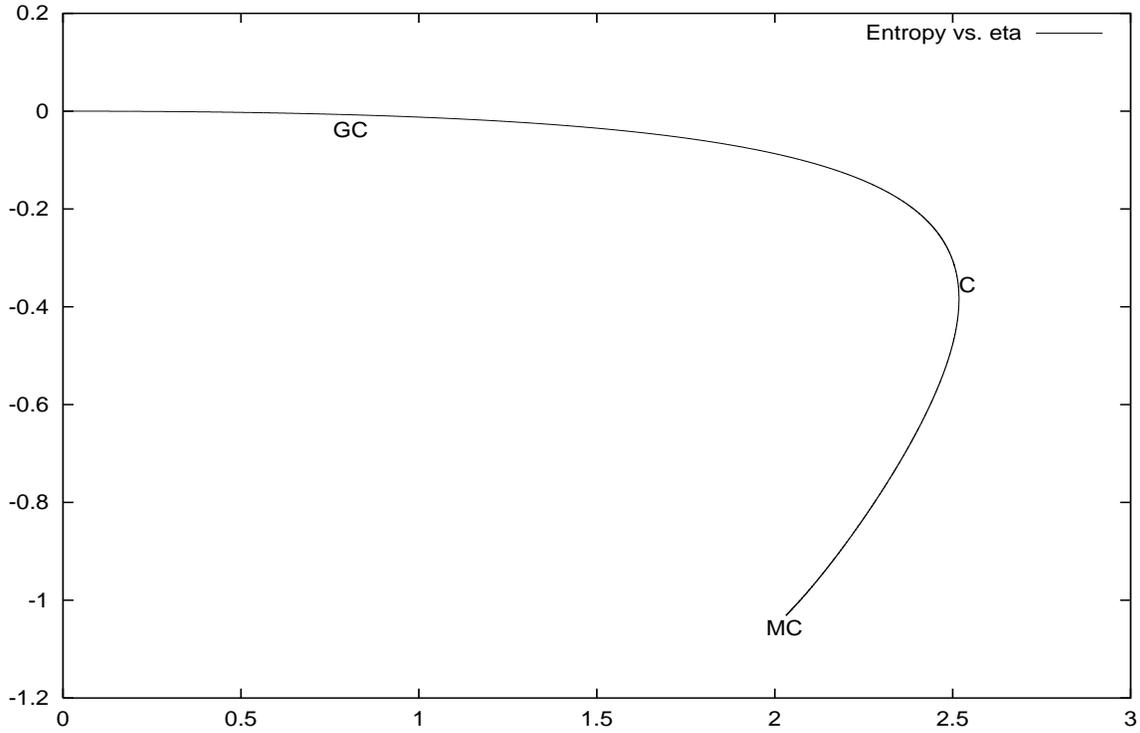,width=10cm,height=16cm} 
\end{turn}
\caption{ The entropy per particle minus the ideal gas value 
 as a function of $\eta^R$ in the MF approximation [eq.(\ref{abel2})].
\label{fig8}} 
\end{figure} 

\begin{figure}
\begin{turn}{-90}
\epsfig{file=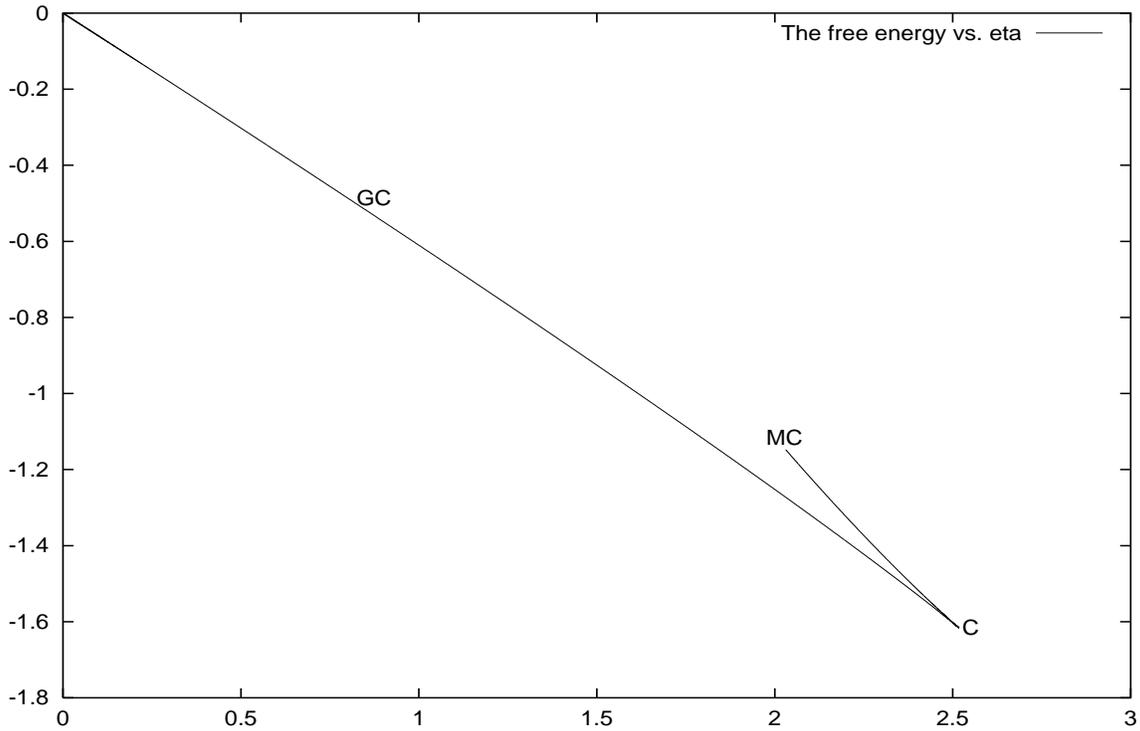,width=10cm,height=16cm} 
\end{turn}
\caption{ $ {F - F_0 \over NT} $ as a function of $ \eta $ in the MF
approximation [eq.(\ref{abel2})]. \label{fig9}} 
\end{figure} 

where 
\begin{equation}\label{fetexpl}
f_{MF}(\eta^R) = {\lambda^2 \over 3 \, \eta^R}\; e^{\chi(\lambda)}
\quad , 
\end{equation}
The function $ f_{MF}(\eta^R) $ obeys the Abel equation,
\begin{equation}\label{abel}
\eta^R(3f_{MF}-1)f'_{MF}(\eta^R)+(3f_{MF}-3+\eta^R) f_{MF} = 0 \;.
\end{equation}
For $ \eta^R = 0 $ it follows from eq.(\ref{abel}) that
\begin{equation}\label{eta0}
 f_{MF}(0) = 1 \; .
\end{equation}
Integrating eq.(\ref{abel}) with respect to $ \eta^R $ yields,
$$
3 \int_0^{\eta^R} {dx \over x} [1 - f_{MF}(x)] = 3 [f_{MF}(\eta^R) - 1
] + \eta^R - \log f_{MF}(\eta^R)
$$
We derive in \cite{I} the properties of the function $ f_{MF}(\eta^R)
$ from the differential equation (\ref{abel}). One easily obtains for
small $ \eta^R $ (dilute regime),
$$
f_{MF}(\eta^R) = 1 - \frac{\eta^R}5 - \frac{(\eta^R)^2}{175} + {\cal
O}([\eta^R]^3)\; . 
$$
These terms exactly coincide with the perturbative calculation in the
dilute regime for spherical symmetry \cite{I}.

We plot in fig. \ref{fig5} $ f_{MF}(\eta^R) $ as a function of
$\eta^R$ obtained by solving eq.(\ref{abel}) by the Runge-Kutta
method. We see that $ 
f_{MF}(\eta^R) $ is a {\bf monotonically decreasing} function of $
\eta^R $ for $ 0 < \eta^R < \eta^R_C $. At the point $ \eta^R =
\eta^R_C $, the derivative $ f_{MF}'(\eta^R) $ takes the value $
-\infty $. It then follows from eq.(\ref{abel}) that 
$$
f_{MF}(\eta^R_C) = \frac13 \; .
$$
At the point $ \eta^R_C $ the series expansion for $ f_{MF}(\eta^R) $ in
powers of $  \eta^R $ diverges. Both, from the ratio test on its
coefficients and from the Runge-Kutta solution, we find that
\begin{equation}\label{etacr}
\eta^R_C = 2.517551\ldots 
\end{equation}
From eq.(\ref{abel}) we find that $ f_{MF}(\eta^R) - \frac13 $ has a
square root behaviour around $ \eta^R = \eta^R_C $:
$$
f_{MF}(\eta^R) \buildrel{ \eta^R \uparrow \eta^R_C}\over= \frac13 +
\sqrt{2(\eta^R_C-2) \over 9 \, \eta^R_C} \; \sqrt{\eta^R_C-\eta^R} 
+{2 \, (\eta^R_C-1) \over 7 \, \eta^R_C} \; (\eta^R_C-\eta^R)
+ {\cal O}\left[(\eta^R_C-\eta^R)^{3/2}\right]
$$
Inserting the numerical value (\ref{etacr}) for $ \eta^R_C $ yields,
\begin{equation}\label{cercaetac}
f_{MF}(\eta^R) \buildrel{ \eta^R \uparrow \eta^R_C}\over= \frac13 +
0.213738\ldots \sqrt{\eta^R_C-\eta^R} +0.172225\ldots\; (\eta^R_C-\eta^R)
+ {\cal O}\left[(\eta^R_C-\eta^R)^{3/2}\right]
\end{equation}
We see that $ f_{MF}(\eta^R) $ becomes complex for $ \eta^R > \eta^R_C
$. Recall that in the Monte Carlo simulations the  gas phase collapses at
the point $ \eta^R_T < \eta^R_C $. $ f_{MF}(\eta^R) $ is a multivalued
function of $ \eta^R $ as well as all physical magnitudes [see
  eq.(\ref{abel2})]. 

The points GC, C and MC correspond to the collapse phase
transition in the grand canonical, canonical and microcanonical
ensembles, respectively. Their positions are determined by the
breakdown of the mean field approximation  through the analysis of the
small fluctuations. 

As noticed before, the CE  only describes  the region between the
ideal gas point, $\eta^R = 0$ and $C$ in fig. 1. The MCE goes beyond 
the point $C$ (till the point $MC$) with the physical 
magnitudes described by the second sheet of the square root in
eqs.(\ref{cercaetac}) (minus sign). We have near $C$ between $C$ and $MC$,
\begin{eqnarray}
&& f_{MF}(\eta^R) \buildrel{ \eta^R \uparrow \eta^R_C}\over= \frac13 -
0.213738\ldots \sqrt{\eta^R_C-\eta^R} + 0.172225\ldots\; (\eta^R_C-\eta^R)
+ {\cal O}\; \left[ (\eta^R_C-\eta^R)^{3/2}\right] \nonumber
\end{eqnarray}

The function $ f_{MF}(\eta^R) $ takes its absolute minimum at $ \eta^R
= \eta^R_{min} = 2.20731\ldots $ in the second sheet where  $
f_{MF}(\eta^R_{min}) = 0.264230\ldots$. 
Since $ f_{MF}(\eta^R) < \frac12 $ implies that the total energy is
negative [see eq.(\ref{abel2})], the gas is in a {\it `bounded state'}
for  $ \eta^R $ beyond  $ \eta^R_2 = 2.18348\ldots  $ in the first sheet.

\bigskip

\begin{figure}[t] 
\begin{turn}{-90}
\epsfig{file=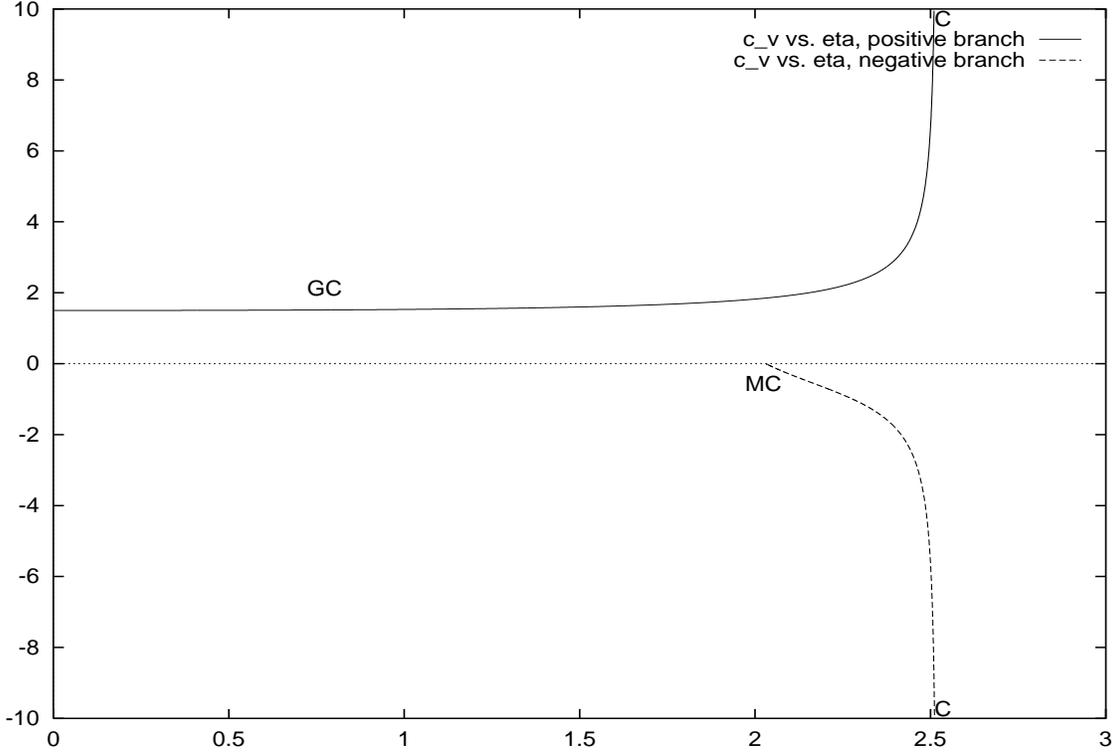,width=10cm,height=16cm} 
\end{turn}
\caption{ $ (c_V)_{MF} $ as a function of $ \eta^R $ from mean field
eq.(\ref{cvmf}). Notice that $ (c_V)_{MF} $ diverges at the point C, that is
for $ \eta^R_C = 2.517551\ldots $ \label{fig1}} 
\end{figure}

\bigskip

For the main physical magnitudes in the mean field approach,
[that is, from the above saddle point and neglecting the fluctuations
around it] we find:
\begin{eqnarray}\label{abel2}
{p V \over NT} &=& f_{MF}(\eta^R) \cr \cr
{F - F_0 \over NT}  &=& 3 [1 - f_{MF}(\eta^R)] - \eta^R + \log f_{MF}(\eta^R)
\cr \cr
{S - S_0 \over N} &=& 6 [f_{MF}(\eta^R)-1] + \eta^R - \log f_{MF}(\eta^R)
\\ \cr
{E \over NT} &=& 3 [f_{MF}(\eta^R)-\frac12] \; , \nonumber 
\end{eqnarray}

We find for the speed of sound squared at the surface 
\begin{equation}\label{vsmf}
{v_s^2 \over T} = {  f_{MF}(\eta^R) \over 3 }\, \left[ 4 + { 3\,
f_{MF}(\eta^R) + \frac{\eta^R}2 - 2 \over 6 \, f_{MF}^2(\eta^R)+
\left(\eta^R - \frac{11}2 \right) f_{MF}(\eta^R) + \frac12 } \right]  \;,
\end{equation}
The specific heat at constant volume takes the form
\begin{equation}\label{cvmf}
(c_V)_{MF}= 6 \, f_{MF}(\eta^R)- \frac72 + \eta^R + {\eta^R - 2 \over
3 \, \, f_{MF}(\eta^R) - 1} \; .
\end{equation}
We plot in Fig. \ref{fig1} eq.(\ref{cvmf}) for $ (c_V)_{MF} $ as a function of
$ \eta $. We see that $ (c_V)_{MF} $ increases with $ \eta $ till it tends to
$ + \infty $ for  $ \eta^R \uparrow \eta^R_C $. It has a square-root
branch point at the point C. In the stretch C-MC (only  physically
realized in the 
microcanonical ensemble), $ (c_V)_{MF} $ becomes negative. We shall
not discuss here the peculiar properties of systems with negative $
C_V $ as they can be find in refs.\cite{lynbell,lynbell2,bt}

From eqs.(\ref{cercaetac}) and (\ref{cvmf}) we obtain the following behaviour
near the point $C$ in the positive (first) branch
\begin{eqnarray}\label{mfcri}
(c_V)_{MF}&\buildrel{ \eta^R \uparrow \eta^R_C}\over=& 0.80714\ldots
(\eta^R_C-\eta^R)^{-1/2} - 0.19924\ldots+ {\cal O}(\sqrt{\eta^R_C-\eta^R})
\end{eqnarray}
and between $C$ and $MC$ in the negative (second) branch 
\begin{eqnarray}
(c_V)_{MF}&\buildrel{ \eta^R \uparrow \eta^R_C}\over=&
-0.80714\ldots(\eta^R_C-\eta^R)^{-1/2} - 0.19924\ldots+{\cal
O}(\sqrt{\eta^R_C-\eta^R})\nonumber 
\end{eqnarray}
Finally,  $ (c_V)_{MF} $ vanishes at the point MC $ \eta^R_{MC}=
2.03085\ldots $. 

\begin{figure}[t] 
\begin{turn}{-90}
\epsfig{file=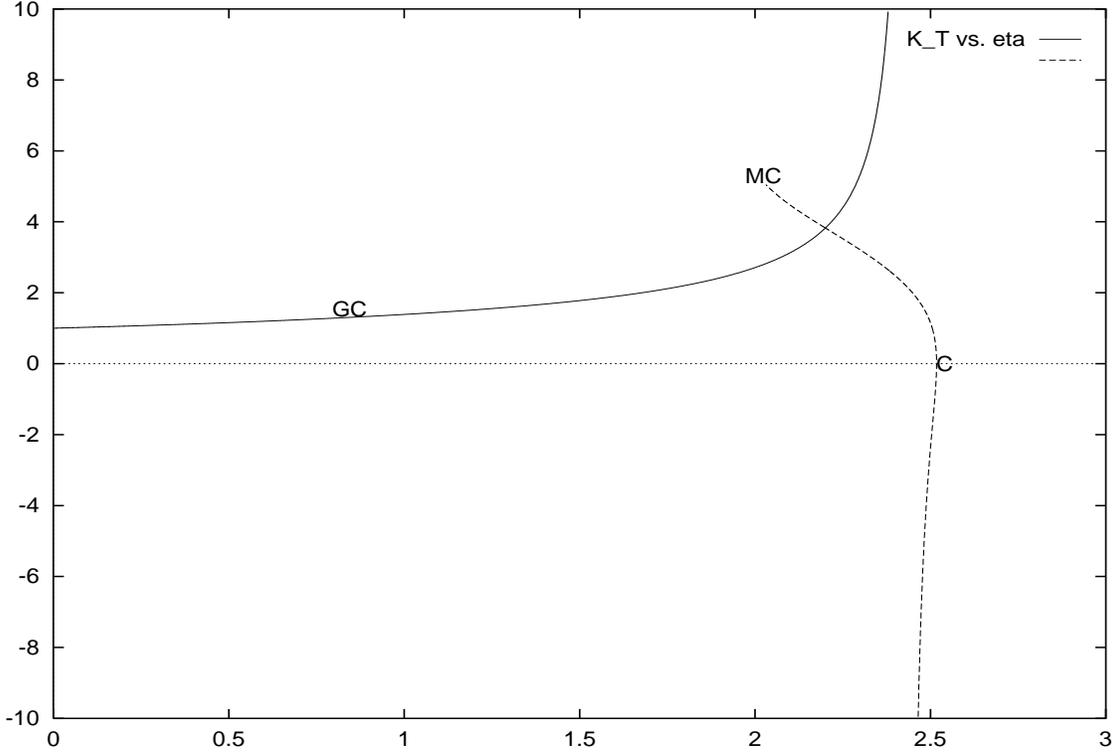,width=10cm,height=16cm} 
\end{turn}
\caption{ $ (\kappa_T)_{MF} $ as a function of $ \eta $ from mean field
eq.(\ref{ktmf}). Notice that $ (\kappa_T)_{MF} $ diverges at $ \eta^R
= \eta_T^R = 2.43450\ldots $ \label{fig11}}  
\end{figure}

The isothermal compressibility in mean field follows from
eqs.(\ref{KT}) and (\ref{abel})
\begin{equation}\label{ktmf}
(\kappa_T)_{MF} = {3 \over 2  f_{MF}(\eta^R)} \, \left[ 1 + {
\eta^R - 2 \over 6  f_{MF}(\eta^R) - \eta^R}\right]\; .
\end{equation}
We plot $ (\kappa_T)_{MF} $ in fig. \ref{fig11}. We see that
$ (\kappa_T)_{MF} $ is positive for $ 0 \leq \eta^R < \eta^R_T =
 2.43450\ldots$ where $ (\kappa_T)_{MF} $ diverges. The point $ \eta^R_T $ 
is defined by the equation
\begin{equation}\label{defeta0}
6  f_{MF}(\eta^R_T) - \eta^R_T = 0\; .
\end{equation}
We find from eqs.(\ref{abel}) and (\ref{defeta0}) that
\begin{equation}\label{fpeta0}
 f_{MF}'(\eta^R_T) = - \frac12
\end{equation}
$ (\kappa_T)_{MF} $ diverges for $ \eta^R \simeq \eta^R_T $ as
$$
(\kappa_T)_{MF} \buildrel{\eta^R \simeq \eta^R_T }\over=
{9 \, (\eta^R_T-2) \over 4 \, \eta^R_T (\eta^R_T - \eta^R)} + {\cal O}(1) = {
0.40157\ldots  \over \eta^R_T - \eta^R} + {\cal O}(1) \; .
$$
$ (\kappa_T)_{MF} $ is  negative for $ \eta^R_T < \eta^R < \eta^R_C $
and exactly  vanishes at the point $ C $. $(\kappa_T)_{MF}$ then
becomes positive in the stretch between $C$ and $MC$ only  physically
realized in the microcanonical ensemble.   

Notice that the singularity of  $(\kappa_T)_{MF}$ at  $ \eta^R =
\eta^R_T =  2.43450\ldots $ is before but near the  point $C$. It 
appears as a preliminary signal of the phase transition at
$C$. $\eta^R_T$ is the transition point $ \eta_T $ seen with the Monte
Carlo simulations (see fig. 1). (Recall that $  \eta_T \sim 1.515 $
corresponds to  $ \eta^R_T \sim 2.44 $).

\bigskip

It is easy to understand the meaning of a large compressibility. From the
definition (\ref{KT})
\begin{equation}\label{compre}
{\delta V \over V} = - K_T \; \delta p = - \kappa_T \; {V \; \delta p
\over N \, T} \; .
\end{equation}
A large compressibility implies that a small increase in the pressure
($ \delta p \ll NT/V $) produces a large change in the
density of the gas. That means a very soft fluid. 

For {\bf negative} compressibility, eq.(\ref{compre}) tells us that the gas
{\bf increases} its volume when the external pressure on it
increases. This is clearly an unusual behaviour that leads to
instabilities as one sees by computing the velocity of sound as a
function of $r$\cite{I,II}.

For the specific heat at constant pressure we find 
\begin{equation}\label{cpmf}
(c_P)_{MF}= 12 \, f_{MF}(\eta^R)- \frac32 + {24\left(\eta^R -
2\right) f_{MF}(\eta^R) \over 6 \, \, f_{MF}(\eta^R) - \eta^R}
\end{equation}
In Table 1 we summarize the critical points and their main properties
for the three ensembles.

\vspace{1cm}
\begin{turn}{90}
\begin{tabular}{|l|l|l|l|l|l|}\hline
$$ & $$ & $$ & $$ & $$ & \\
 POINT  & $\hspace{0.8cm}\lambda$ & $\hspace{0.8cm}\eta^R$ &
\hspace{2cm} Defining Equation & $ \hspace{0.2cm} f_{MF}(\eta^R) $ &
PHYSICAL MEANING \\ 
$$ & $$ & $$ & $$ & $$ &\\ \hline 
$$ & $$ & $$ & $$ & $$ &\\
$\;$ GC  &   $ 1.7772\ldots $ & $0.797375\ldots $ & $\hspace{1.75cm} 2 - 3\;
\eta^R_{GC}\; f_{MF}(\eta^R_{GC}) = 0 $ & $ 0.836076\ldots $ & Collapse
in the GCE. \\ 
$$ & $$ & $$ & $$ & $$ & $$ \\ \hline 
$$ & $$ & $$ & $$ & $$ &  Energy density \\ 
$\; \; \; \; 3 $ & $ 3.38626\ldots $ & $
1.73745\ldots     $ & $\hspace{1.75cm} 3 - \eta^R + \chi(\lambda) = 0 
$ & $ 0.622424\ldots  $ & vanishes at $r=0$. \\
$$ & $$ & $$ & $$ & $$ & $$ \\ \hline 
$$ & $$ & $$ & $$ & $$ & $$ \\
$\; \; \; \;  2$ & $4.73739\ldots $ & $ 2.18348\ldots$ &\hspace{2.1cm} $2\;
f_{MF}(\eta^R_2) - 1 = 0 $ & $\hspace{0.5cm} 0.5$ & Total Energy vanishes.\\ 
$$ & $$ & $$ & $$ & $$ &$$ \\ \hline
$$ & $$ & $$ & $$ & $$ &$$ \\
$ \; \; \; \; T$ & $6.45077\ldots  $ & $ 2.43450\ldots$ &\hspace{2.1cm} $6\;
f_{MF}(\eta^R_T) - \eta^R_T = 0 $ & $0.40575\ldots$ & $\kappa_T$ and
$c_P$ diverge. \\ 
$$ & $$ & $$ & $$ & $$ & Collapse in the CE.\\ \hline 
$$ & $$ & $$ & $$ & $$ & $$ \\
$ \; \; \;  $ C &  $ 8.993195\ldots  $ &  $ 2.517551\ldots  $ &
$\hspace{2.1cm} 1 
- 3 \; f_{MF}(\eta^R_{C}) = 0 $ & $ \hspace{0.5cm} 1/3 $ & 
$ c_V $ diverges. \\ 
$$ & $$ & $$ & $$ & $$ &$$ \\\hline
$$ & $$ & $$ & $$ & $$ & Minimum of\\
$ \; \; $ Min & $22.5442\ldots $ & $2.20731\ldots$ &\hspace{2.1cm} $
f'_{MF}(\eta^R_{min}) = 0 $ &  $ 0.264230\ldots$ & pV/[NT] \\ 
$$ & $$ & $$ & $$ & $$ & in the gas phase\\ \hline
$$ & $$ & $$ & $$ & $$ & $$\\
$ \; \; \; \; 1 $ & $25.7991\ldots $ & $ 2.14675\ldots $ & $ \; \;
 48f_{MF}^2(\eta^R_1) - (38 - 8 \eta^R_1)\, f_{MF}(\eta^R_1) +
$ & $0.265290\ldots$ & $v_s^2$ and $c_P$ vanish.
\\ $$ & $$ & $$ &\hspace{3cm} $  + \eta^R_1 = 0  $ & $$ &\\ \hline 
$$ & $$ & $$ & $$ & $$ & Collapse in the MCE. \\
 $ \; \; $ MC  & $ 34.36361\ldots $ & $ 2.03085\ldots $ &
$ 12f_{MF}^2(\eta^R_{MC}) - (11 - 2 \eta^R_{MC})\, f_{MF}(\eta^R_{MC})+
$  & $ 0.273512\ldots $ & $c_V$ vanishes.
\\ $$ & $$ & $$ & \hspace{3cm} $+ 1 = 0$ & $$ & $$ \\ \hline
\end{tabular}
\end{turn}
\vspace{0.5cm}

{TABLE 1. Values of the critical points in the three ensembles GC, C
and MC (using mean field) and further
characteristic points for spherical symmetry. $ pV/[NT], E $ and $ S \to
-\infty $ for $  \eta^R \uparrow \eta^R_{GC} $ and $ \eta^R \uparrow
\eta^R_C $. Notice that $ \eta_{Min}, \; 
\eta_1 $ and $ \eta_{MC} $ are in the second Riemann sheet.}

\subsection{Average distance between particles}

We investigate here the average distance between particles $ <r> $ and
the average squared distance $ <r^2> $. The study of $ <r> $ and $
<r^2> $ as functions of $ \eta $ permit a better understanding of the
self-gravitating gas and its phase transition in the different
statistical ensembles. 

These average distances are defined as
\begin{eqnarray}\label{defryr2}
<r> &\equiv& \int \int\left| {\vec r}- {\vec r \,}' \right| \; <
\rho({\vec r}) \; 
\rho({\vec r \, }') > \; d^3 r \; d^3 r'  \; , \cr \cr
<r^2> &\equiv& \int \int\left| {\vec r}- {\vec r \, }' \right|^2  \; <
\rho({\vec r}) \; \rho({\vec r \, }') > \; d^3 r \; d^3 r'
\end{eqnarray}
In the mean field approximation we have 
\begin{equation}\label{correMF}
< \rho({\vec r}) \; \rho({\vec r \, }') > = \rho_{MF}({\vec r}) \;
\rho_{MF}({\vec r \, }') + {\cal O}\left( {1 \over N} \right) \; .
\end{equation}
In addition, in the spherically symmetric case we use eq.(\ref{densr})
for the particle density. In Appendix E we compute the 
integrals in eqs.(\ref{defryr2}) and we get as result,
\begin{eqnarray}\label{ryr2exp}
<r> &=& 2 - 2 \; \int_0^1 r^2 \; dr \; \left[ 1 + { \phi(r) -\phi(1)
\over \eta^R} \right]^2  \; , \cr \cr
<r^2> &=& 2 - {12 \over \eta^R}\; \int_0^1 r^2 \; dr \; \left[ 
\phi(r) -\phi(1) \right]
\end{eqnarray}
where $\phi(r)$ is given by eq.(\ref{fixi}).

\begin{figure}[t] 
\begin{turn}{-90}
\epsfig{file=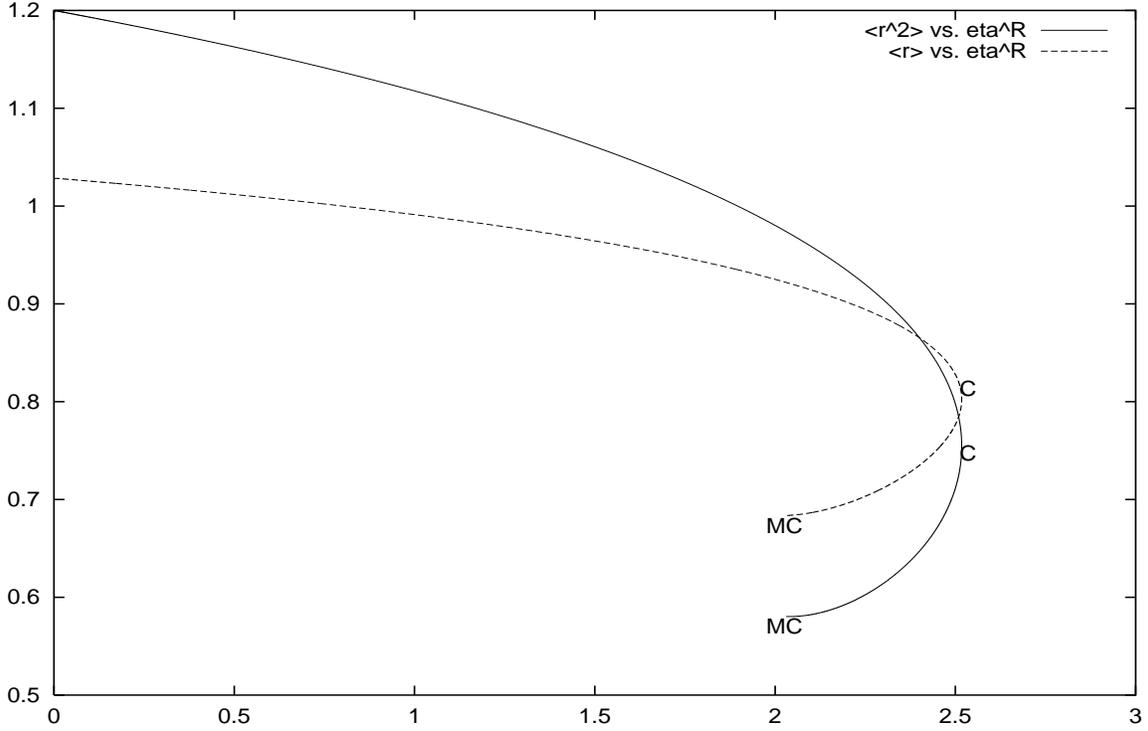,width=10cm,height=16cm} 
\end{turn}
\caption{Mean value of the distance between particles ($ <r> $) and
mean value of the squared distance between particles
($ <r^2> $) as functions of $ \eta^R $ in the mean
field approach from eqs.(\ref{ryr2exp}). Notice that the particles are
inside a sphere of unit radius. 
\label{fig15}} 
\end{figure}

We plot $ \, <r> $ and $ <r^2> $ as functions of $
\eta^R $ in fig. \ref{fig15}. Both $ <r> $ and $ <r^2> $
monotonically decrease with $ \lambda(\eta^R) $. Their values for the
ideal gas are 
$$
\left. <r>\right|_{\eta=0} = \frac{36}{35}=1.02857\ldots \quad , \quad\left.
<r^2>\right|_{\eta=0} = \frac65 \; .
$$
At the critical points (C for the canonical ensemble and MC for the
microcanonical ensemble) the average distances
sharply decrease. Both  $ \, <r> $ and $ <r^2> $ have infinite
slope as functions of $\eta^R $ at the point C. 

\begin{figure}[t] 
\begin{turn}{-90}
\epsfig{file=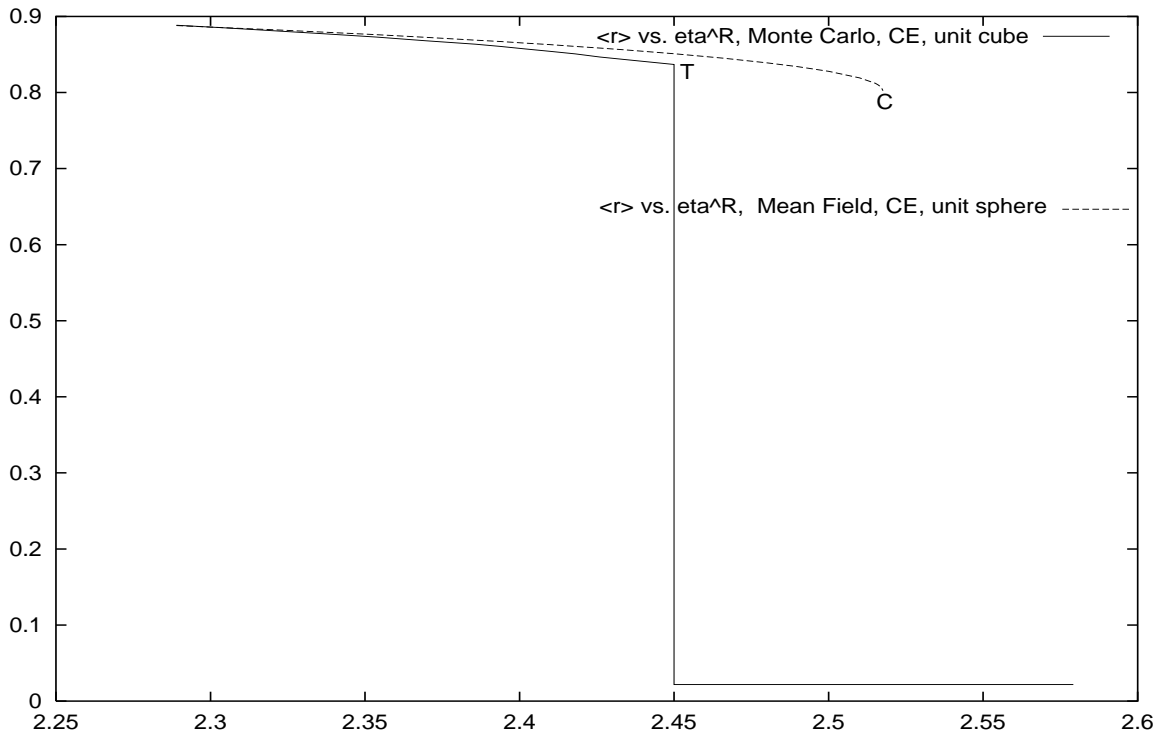,width=10cm,height=16cm} 
\end{turn}
\caption{Mean value of the distance between particles ($ <r> $) in a unit cube
from Monte Carlo simulations and in a unit sphere from mean field as
functions of $ \eta^R $. 
\label{rmmcmf}} 
\end{figure}

We plot in fig. \ref{rmmcmf} the Monte Carlo results for $ <r> $ in a unit cube
together with the MF results in a unit sphere. Notice that $ <r> $
sharply falls at the point T clearly indicating the transition to collapse. 

\subsection{Particle Distribution}

The particle distribution at thermal equilibrium obtained through the Monte 
Carlo simulations and mean field methods is inhomogeneous both in the
gaseous and condensed phases.  In the dilute regime $\eta \ll 1$ the gas
density is uniform, as expected. 

For the Monte Carlo simulations the density of particles in
the cube are shown for the gaseous and for the condensed phases in figs. 3-6,
respectively. 

In the mean field approximation and for the spherically symmetric case,
the particle density is given by
\begin{equation}\label{densr}
\rho_{MF}(r) = e^{\phi(r)} = {\lambda^2 \; e^{\chi(\lambda r)} \over  4\, \pi
\,\eta^R} \quad , \quad 0 \leq r \leq 1\; .
\end{equation}

The mass inside a radius $ r $ is then given by
$$
M(r) = 4\, \pi \int_0^r {r'}^2 \; \rho_{MF}(r')\;dr' = - { \lambda \; r^2
\over\eta^R} \, \chi'(\lambda r) \; ,
$$
where we used eq.(\ref{ecuaxi}). For small $ r $ this gives
\begin{equation}\label{rchico}
M(r)\buildrel{r \ll 1}\over = { \lambda^2 \; r^3 \over3 \, \eta^R}
\left[ 1 + {\cal O}(\lambda^2 \; r^2) \right]\; .
\end{equation}
We find an uniform mass distribution near the origin. This is simply explained
by the absence of gravitational forces at $ r = 0 $. Due to the
spherically symmetry,
the gravitational field exactly vanishes at the origin. The particles
exhibit a perfect gas distribution in the vicinity of $ r = 0
$. Actually, eq.(\ref{rchico}) is {\bf both} a short distance and a
{\bf weak coupling} expression. Eq.(\ref{rchico}) is valid in the
dilute limit $ \eta^R \ll 1 $ for all $ 0 \leq r \leq 1 $.

\begin{figure}[t] 
\begin{turn}{-90}
\epsfig{file=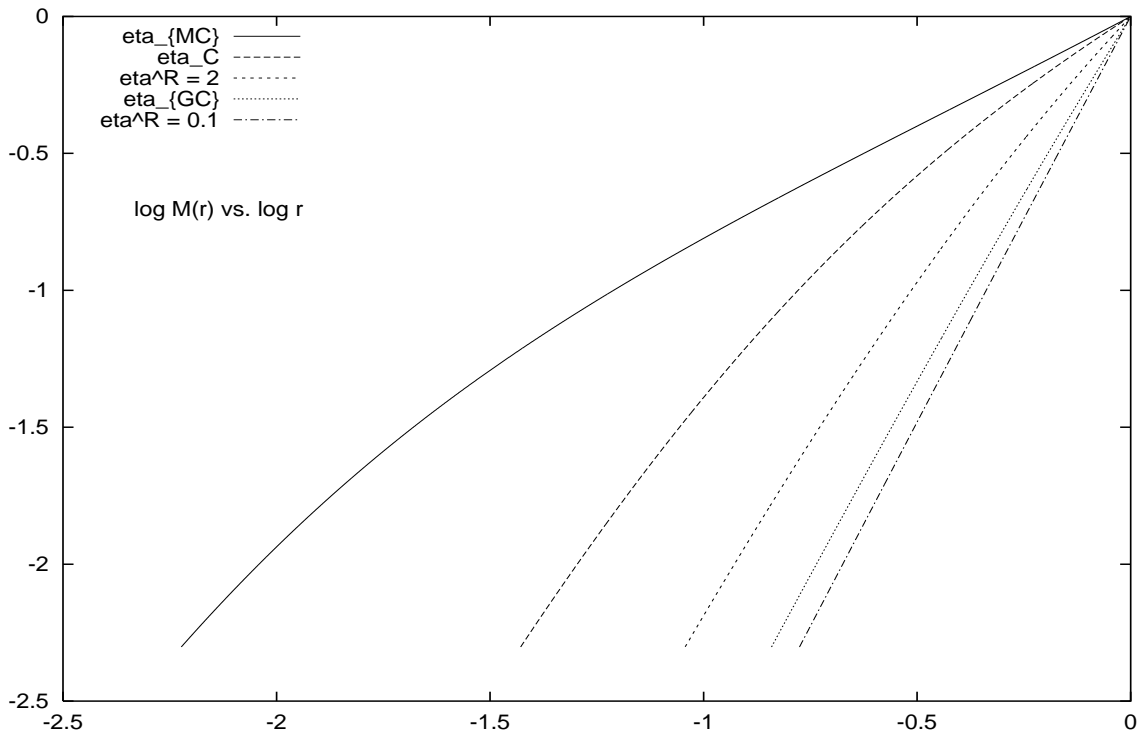,width=10cm,height=16cm} 
\end{turn}
\caption{ $ \log M(r) $ vs. $ \log r $ for five different values of $
\eta^R : \; \eta^R_{MC} = 2.03085\ldots , \; \eta^R_{C} =
2.517551\ldots, \; \eta^R =2 , \; \eta^R_{GC} = 0.797375\ldots $ and
$ \eta^R = 0.1 $. \label{fig12}} 
\end{figure}

\vspace{1cm}

\begin{tabular}{|l|l|l|}\hline
$ \eta^R $ & $\hspace{0.5cm} D $ & $\hspace{0.5cm} C $\\ \hline 
$ 0.1 $ &    \hspace{0.3cm} $ 2.97 $ \hspace{0.3cm}  &  \hspace{0.3cm}
$ 1.0 $ \hspace{0.3cm}  \\ \hline 
$ \eta^R_{GC} = 0.797375\ldots $ & \hspace{0.3cm} $ 2.75  $\hspace{0.3cm} &
\hspace{0.3cm} $ 1.03 $ \hspace{0.3cm} \\ \hline 
 $ 2.0 $  & \hspace{0.3cm} $ 2.22 $ \hspace{0.3cm} &\hspace{0.3cm} $ 1.1
$\hspace{0.3cm} \\ \hline 
$ \eta^R_{C} = 2.517551\ldots $ & \hspace{0.3cm} $ 1.60
$\hspace{0.3cm} & \hspace{0.3cm} 
$ 1.07 \hspace{0.3cm} $ \\ \hline
$ \eta^R_{MC} = 2.03085\ldots $ & \hspace{0.3cm} $ 0.98  $
\hspace{0.3cm}&\hspace{0.3cm} 
$ 1.11 $\hspace{0.3cm} \\ \hline 
\end{tabular}

\vspace{0.5cm}

{TABLE 2. The Fractal Dimension $ D $ and the proportionality
coefficient $ C $ as a function of $ \eta^R $ from a fit to the mean
field results according to $ { \cal M }(r) \simeq C \; r^D $.}

\vspace{0.5cm}

We plot in fig. \ref{fig12} the particle distribution for $ 90 \% $ of
the particles for several values of $ \eta^R $. We exclude in the
plots the region $ M(r) < 0.1 $ where the distribution is uniform.

We find that these mass distributions approximately follow the power law
\begin{equation}\label{esca}
{ \cal M }(r) \simeq C \; r^D 
\end{equation}
where, as depicted in Table 2, $ D $ slowly decreases with  $
\lambda(\eta^R) $ from the value $ D = 3 $ for the ideal gas ($\eta=0$) till $
D = 0.98 $ in the extreme limit of the MC point.

\subsection{The speed of sound as a function of  ${\vec r}$}

For very short
wavelengths $ \lambda_s \ll L $, the sound waves just feel the local
equation of state (\ref{pr}) and the speed of sound will be that of an
ideal gas. For long wavelengths (of the order $L$), the situation
changes. The calculation in eq.(\ref{vsmf}) corresponds to the speed of sound
for an external wave arriving on the sphere in the long wavelength limit. 
Let us now make the analogous calculation for a wave reaching the
point $ {\vec q} $ inside the gas.

Our starting point is eq.(\ref{defson}),
\begin{equation}\label{vsr}
v_s^2({\vec q}) = - {{ c_P \; V^2} \over {c_V \; N}} \left({ \partial
p({\vec q}) \over \partial V}\right)_{T,{\vec q}} \; ,
\end{equation}
where $ c_P $ and $ c_V $ are the specific heats of the whole system
at constant (external) pressure and volume, respectively, and $
p({\vec q}) $ is the local pressure at the point $ {\vec q} $.
The local pressure in the spherically symmetric case can be written in
a more explicit way using eqs.(\ref{fixi}) and (\ref{pr}):
\begin{equation}\label{prad}
{ p(r) \; V \over N \, T} = {\lambda^2 \over 3 \, \eta^R}\;
e^{\chi(\lambda\, r)}
\end{equation}

We find for the spherically symmetrical case in MF
\begin{equation}\label{vsre}
{v_s^2(r) \over T} = { c_P \over c_V} \; { \lambda^2 \over 9 \, \eta^R
\; \left[3 f(\eta^R) -1\right] } \; \left[ 6 \, f(\eta^R) + \lambda \, r \;
\chi'(\lambda\, r)\right] \; e^{\chi(\lambda\, r)}\; ,
\end{equation}
where we used eqs.(\ref{prad}), (\ref{vsr}) and
$$
\left({\partial \eta \over \partial V}\right)_T = - {\eta \over 3 \, V} \; .
$$
[$\lambda$ is a function of $\eta^R$ as defined by eq.(\ref{lambaxi})].

At the surface, $ (r=1) , \;  v_s^2(r) $ reduces to eq.(\ref{vsmf}) after using
eqs.(\ref{lambaxi}), (\ref{fetexpl}), (\ref{cvmf}) and (\ref{cpmf}).

\begin{figure}
\begin{turn}{-90}
\epsfig{file=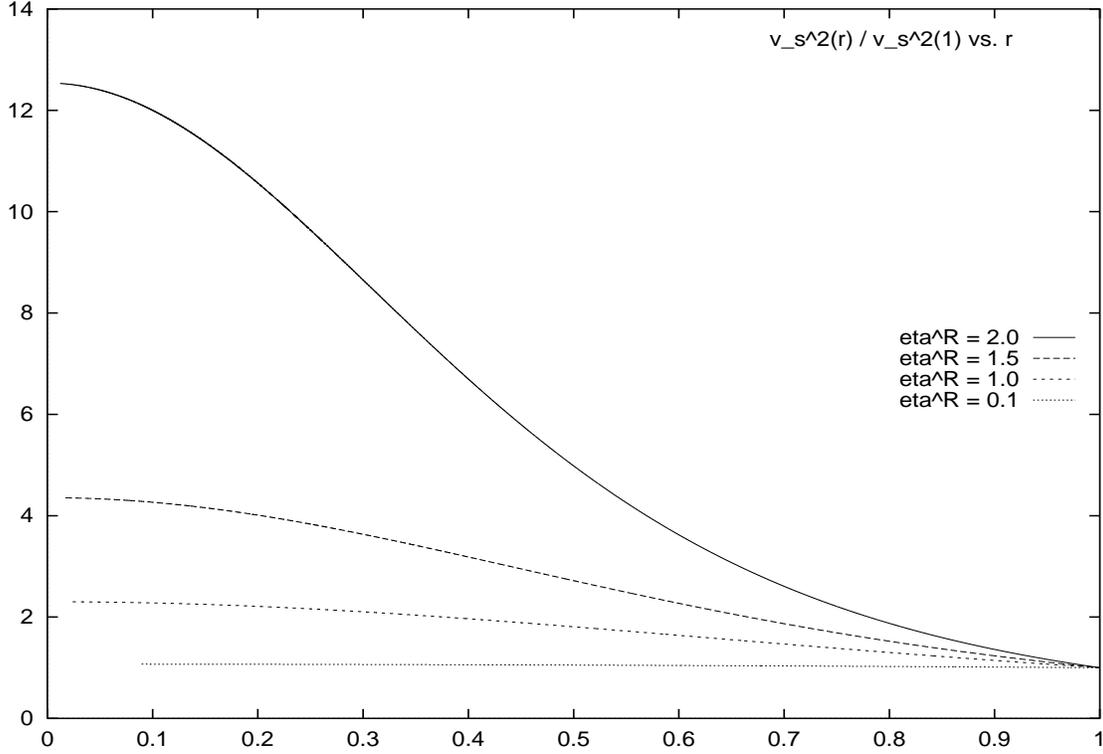,width=10cm,height=16cm} 
\end{turn}
\caption{ The speed of sound $ v_s^2(r) / v_s^2(1)  $ as a function of
  $ r $  for 
$\eta^R= 2.0, \; 1.5, \; 1.0 $ and $ 0.1 $. That is, 
values of $\eta^R$ smaller than  $\eta^R_T=2.43450\ldots$. $ v_s^2(r) /
v_s^2(1)  $ is here always positive and decreases with $ r $.
\label{vsonrinf}}
\end{figure}

\begin{figure}
\begin{turn}{-90}
\epsfig{file=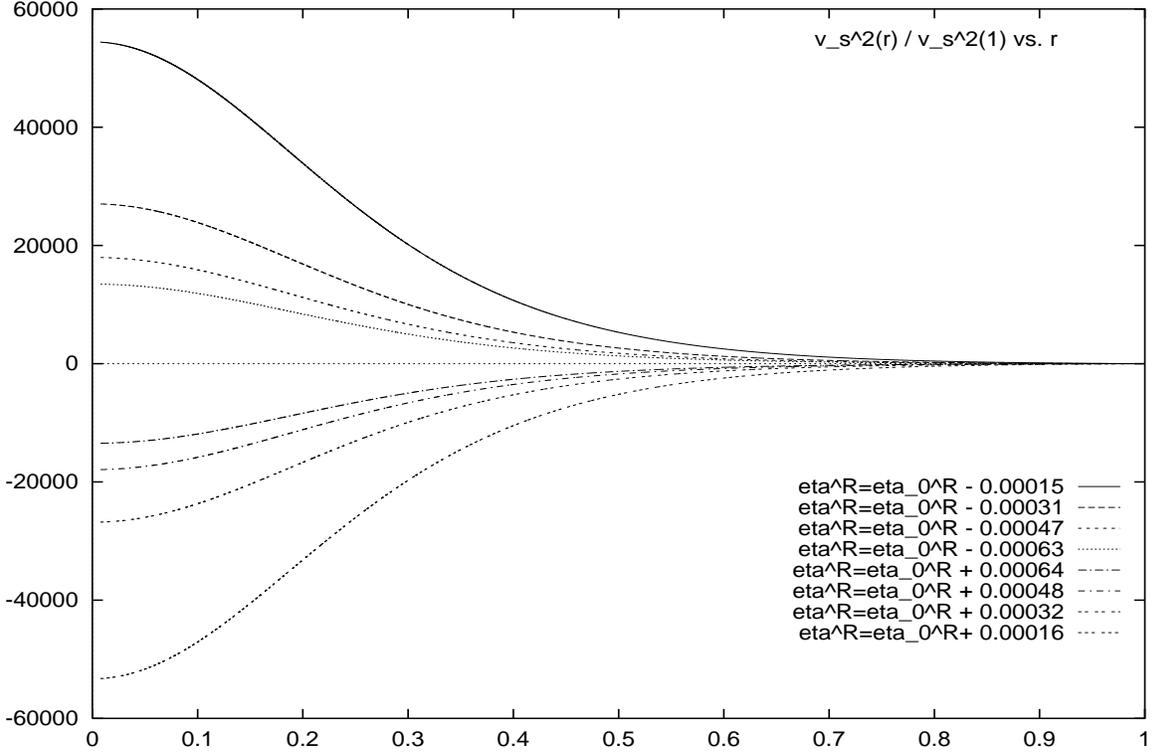,width=10cm,height=16cm} 
\end{turn}
\caption{ The speed of sound  $ v_s^2(r) / v_s^2(1)  $ as a function
  of $ r $  for 
values of $\eta^R$ around $\eta^R_T=2.43450\ldots$. Positive values of
$ v_s^2(r)/ v_s^2(1)  $ correspond to $\eta^R < \eta^R_T$ and negative
values of $ v_s^2(r)/ v_s^2(1)  $ correspond to $\eta^R >
\eta^R_T$. We see that the speed of sound squared tends to $ +\infty $
in the bulk ($ r < 1 $) for $ \eta^R \uparrow \eta^R_T $ while it 
tends to $ -\infty $ for $ \eta^R \downarrow \eta^R_T $.
\label{vsonr}}
\end{figure}

For $ \eta^R < \eta^R_T = 2.43450\ldots $ in the first sheet, $
v_s^2(r) $ is positive and decreases with $ r $ as shown in
fig. \ref{vsonrinf}.  

At $ \eta^R
= \eta^R_T , \; v_s^2(r) $ diverges for all $ 0 \leq r <1 $ due to the factor $
c_P $ in eq.(\ref{vsre}) [cfr. eq.(\ref{cpmf})]. The
derivative of $p$ with respect to $V$ is proportional to $ 6 \,
f(\eta^R) + \lambda \, r \; \chi'(\lambda\, r) $ as we see in
eq.(\ref{vsre}). At $ r=1 $ this 
factor becomes $ 6 \, f(\eta^R) - \eta^R $ [see eq.(\ref{lambaxi})]
which exactly vanishes at $ \eta^R = \eta^R_T $ canceling the
singularity that $ c_P $ possesses at such point [see eq.(\ref{cpmf})].  
Thus, $ v_s^2(1) $ is regular at $ \eta^R = \eta^R_T $.  

$ v_s^2(r) $ becomes {\bf large and positive} below and near $
\eta^R_T $ and {\bf large and negative} above and near $ \eta^R_T =
2.43450\ldots  $ as
depicted in fig. \ref{vsonr}. This singular behaviour witness the
appearance of strong instabilities at $ \eta^R = \eta^R_T $. For 
$ \eta^R > \eta^R_T , \; v_s(r) $ becomes imaginary indicating the
exponential growth of disturbances in the gas. This phenomenon is
especially dramatic in the denser regions (the core). 

For  $ \eta^R $ beyond $ \eta^R_T $ and before $ \eta^R_1 =
2.14675\ldots $ in the second sheet $ v_s^2(r) $ stays negative around
the core while it becomes positive in the external regions 
For example, $ v_s^2(r) $ is positive at $ \eta^R =
\eta^R_C $ for $ r > 0.4526\ldots $. 

For  $ \eta^R $ in the second sheet beyond $ \eta^R_1 $ and before $
\eta^R_{MC} , \;  v_s^2(r) $ is positive in the core and negative
outside. 

\section{$\nu$-dimensional generalization}

The  self-gravitating gas can be studied in $\nu$-dimensional space where
the Hamiltonian takes the form
\begin{equation}\label{hamiD}
H_N = \sum_{l=1}^N\;{{p_l^2}\over{2m}} - G \, m^2 \sum_{1\leq l < j\leq N}
{1 \over { |{\vec q}_l - {\vec q}_j|_A^{\nu-2}}},\quad  {\rm for}\;  \nu \neq 2
\end{equation}
and
\begin{equation}\label{hami2}
H_N = \sum_{l=1}^N\;{{p_l^2}\over{2m}} - G \, m^2 \sum_{1\leq l < j\leq N}
\log{1 \over { |{\vec q}_l - {\vec q}_j|_A}}, \quad  {\rm at}\;  \nu= 2\; .
\end{equation}

The partition function in the microcanonical, canonical and grand
canonical ensembles takes forms analogous to eqs.(II.2),
(III.1) and (VI.7) in paper I, respectively.  

We now find for the microcanonical ensemble,
$$
S(E,N)  = \log\left[ {N^{\nu N-2} \, m^{3\nu N/2-2} \,
L^{\nu(2-\nu/2)N +\nu -2} \, G^{\nu N/2 -1}
\over N !\, \Gamma\left( {\nu N \over 2} \right)\, {(2\pi)}^{\nu N/2}}\right]
+ \log w(\xi,N) \; .
$$
where the coordinate partition function takes now the form
$$
w(\xi,N)\equiv \int_0^1\ldots \int_0^1 \prod_{l=1}^N\; d^\nu r_l \; 
\left[\xi + {1 \over N}u({\vec r}_1, \ldots, {\vec r}_N) \right]^{\nu N/2
-1}\theta\left[\xi + 
{1 \over N}u({\vec r}_1, \ldots, {\vec r}_N) \right]\; .
$$
with
\begin{equation}\label{tzinu}
\xi = { E \, L^{\nu-2} \over G \, m^2 \, N^2}
\end{equation}
and
$$
u({\vec r}_1, \ldots, {\vec r}_N) \equiv {1 \over
N}\sum_{1\leq l < j\leq N} {1 \over { |{\vec r}_l - {\vec r}_j|_a^{\nu-2}}}\; .
$$

In the canonical ensemble we obtain now,
$$
{\cal Z}_C(N,T) = {1 \over N !}\left({m T
L^2\over{2\pi}}\right)^{\frac{\nu N}2}
\; \int_0^1\ldots \int_0^1
\prod_{l=1}^N d^\nu r_l\;\; e^{ \eta \; u({\vec r}_1,\ldots,{\vec r}_N)}
$$
where the variable $ \eta $ takes the form
\begin{equation}\label{etanu}
\eta = {G \, m^2 N \over L^{\nu-2} \; T} \; .
\end{equation}
As we can see from eqs.(\ref{tzinu})-(\ref{etanu}) the only change
going off three dimensional space is in the exponent of $L$.

\bigskip

In $ \nu$-dimensional space the thermodynamic limit is 
defined as $ V, \; N  \to \infty $ keeping  $ \eta $ and $ \xi $
fixed. That is, $ N/ L^{\nu-2} = N /
V^{1-2/\nu} $ is kept fixed. The volume density of particles $ N/V $
vanishes as  $ V^{-2/\nu} $ for $ V, \; N  \to \infty $ and $ \nu > 2
$. It is a  dilute limit for $ \nu > 2 $.

When $ \nu \leq  2 $, one has to assume that the temperature tends to
infinity in the thermodynamic limit in order to keep $ \eta $ and $ \xi $
fixed as $ V, \; N  \to \infty $. 

\section{Functional Determinants: the Validity of Mean Field} 

The mean field gives the dominant behaviour for $ N \to \infty $. 
 The Gaussian functional integral of small fluctuations around the
 stationary points was evaluated in ref.\cite{II}.

As remarked in \cite{I}, the three statistical ensembles (grand
canonical, canonical and microcanonical) yield identical results for
the saddle point. However, the small fluctuations around the
saddle take different forms in each ensemble.

\begin{figure}
\begin{turn}{-90}
\epsfig{file=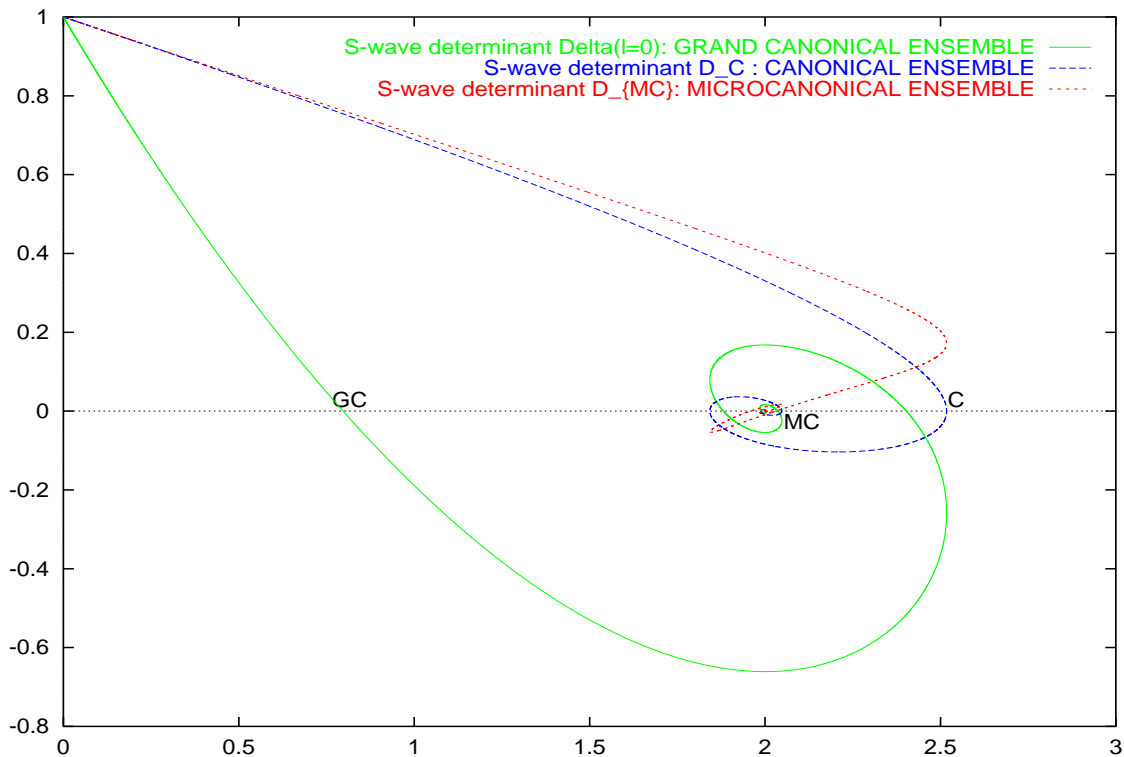,width=10cm,height=16cm} 
\end{turn}
\caption{ 
The S-wave determinants $ D_{GC}(\eta^R), \; D_C(\eta^R) $ and $
D_{MC}(\eta^R) $ in the grand canonical, canonical and
microcanonical ensembles, respectively,  as functions of $ \eta^R
$. Notice that the mean field approximation for  each ensemble breaks
down as soon as the respective determinant becomes negative when $
\eta $ increases starting from $ \eta = 0 $.
\label{fig6}} 
\end{figure}

Adding the contributions from the functional determinant to the mean
field results eq.(\ref{abel2}) in the grand canonical,  canonical and
microcanonical ensembles corresponds to include $ 1/N $ corrections in the
function $ f(\eta^R) $ as follows,
\bea\label{fcano}
&&f(\eta^R) = f_{MF}(\eta^R) + {\eta^R \over 6N}{d \over
d\eta^R}\log D_{GC}(\eta^R) +{\cal O}\left({1 \over
N^2}\right) \; \mbox{grand canonical ensemble}  ; \cr \cr
&&f(\eta^R) = f_{MF}(\eta^R) + {\eta^R \over 6N}{d \over
d\eta^R}\log D_{C}(\eta^R) +{\cal O}\left({1 \over
N^2}\right) \;\mbox{ canonical ensemble}  ; \\ \cr 
&&f(\eta^R) = f_{MF}(\eta^R) + {\eta^R \over 6N}{d \over
d\eta^R}\log D_{MC}(\eta^R) +{\cal O}\left({1 \over
N^2}\right) \;\mbox{ microcanonical ensemble} . \nonumber
\eea
where $D_{GC}, \; D_{C} $ and $ D_{MC} $ are the functional
determinants in the grand canonical, canonical and microcanonical
ensembles, respectively. We plot in fig. \ref{fig6} the S-wave part of
these determinants as functions of $ \eta^R $. The higher partial
waves give a positive definite contribution to these three determinants.

We find in the grand canonical ensemble that
$$
\eta^R{d \over d\eta^R}\log\mbox{Det}_{GC}(\eta^R) \buildrel{ \eta^R \uparrow
\eta^R_{GC}}\over= -{\eta^R_{GC} \over \eta^R_{GC}-\eta^R} \to -\infty \; ,
$$  
Therefore, $ {p V \over NT} $, the energy and the entropy tend to {\bf minus
infinity} when $ \eta^R \uparrow \eta^R_{GC} $. This behaviour
correctly suggests that the  gas collapses for $ \eta^R \uparrow
\eta^R_{GC} $. Indeed, the Monte Carlo simulations yield a large and
negative value for $ {p V \over NT} $ in the collapsed phase \cite{I}.

We want to stress that the mean field values provide {\bf excellent}
approximations as long as $ N|\eta^R_{GC}-\eta^R| >> 1 $ in the grand
canonical ensemble. Namely, the mean field is completely
reliable for large $ N $ unless $\eta^R$ gets at a distance of the
order $ N^{-1} $ from  $\eta^R_{GC}$.

The clumping phase transition in the canonical ensemble takes place
when $D_C(\eta^R)$ vanishes at 
$\eta^R = \eta^R_C$. Near such point the expansion in $1/N$ breaks down
since the correction terms in eq.(\ref{fcano}) become large. 
Mean field applies when $ N|\eta^R_C-\eta^R| >> 1 $.

Since
\begin{equation}\label{fcorcan}
{\eta^R \over 6  }\; { d \over d \eta^R}\log D_C(\eta^R)\buildrel{
\eta^R \uparrow 
\eta^R_C}\over= -{\eta^R_C \over 12(\eta^R_C-\eta^R)} \to -\infty \; ,
\end{equation}
eq.(\ref{fcorcan}) correctly suggests that $PV/[NT], \; E/[NT] $ and
the entropy per particle become {\bf large and negative} for $\eta^R
\uparrow \eta^R_C$. Indeed the Monte Carlo simulations yield a large and
negative value for these three quantities in the collapsed phase\cite{I}.

For $ \eta^R \downarrow\eta^R_{MC} $, reaching the point MC, 
we find
$$
{\eta^R \over 6  }\; { d \over d \eta^R}\log D_C(\eta^R)\buildrel{
\eta^R \downarrow 
\eta^R_{MC}}\over= {\eta^R_{MC} \over 12(\eta^R-\eta^R_{MC})} \to +\infty \; ,
$$
We see that the MF predicts that $ pV/[NT] $ {\bf grows} approaching
the critical point MC. This behaviour is confirmed by the Monte Carlo
simulations. At the point MC, $ pV/[NT] $ increases discontinuously  by
$ 50 \% $ in the Monte Carlo simulations.

\section{The Interstellar Medium}

The interstellar medium (ISM) is a gas essentially formed by atomic (HI) 
and molecular ($H_2$) hydrogen, distributed in cold ($T \sim 5-50 K$) 
clouds, in a very inhomogeneous and fragmented structure. 
These clouds are confined in the galactic plane 
and in particular along the spiral arms. They are distributed in 
a hierarchy of structures, of observed masses from 
$10^{-2} \; M_{\odot}$ to $10^6 M_{\odot}$. The morphology and
kinematics of these structures are traced by radio astronomical 
observations of the HI hyper fine line at the wavelength of 21cm, and of
the rotational lines of the CO molecule (the fundamental line being
at 2.6mm in wavelength), and many other less abundant molecules.
  Structures have been measured directly in emission from
0.01pc to 100pc, and there is some evidence in VLBI (very long based 
interferometry) HI absorption of structures as low as $10^{-4}\; pc = 20$ AU 
(3 $10^{14}\; cm$). The mean density of structures is roughly inversely
proportional to their sizes, and vary 
between $10$ and $10^{5} \; atoms/cm^3$ (significantly above the 
mean density of the ISM which is about 
$0.1 \; atoms/cm^3$ or $1.6 \; 10^{-25}\; g/cm^3$ ).
Observations of the ISM revealed remarkable relations between the mass, 
the radius and velocity dispersion of the various regions, as first 
noticed by Larson \cite{larson}, and  since then confirmed by many other 
independent observations (see for example ref.\cite{obser}). 
From a compilation of well established samples of data for many different  
types of molecular clouds of maximum linear dimension (size) $ R $,  
total mass $M$ and internal velocity dispersion $ \Delta v$ in each region: 
\begin{equation}\label{vobser}
M (R)  \sim    R^{d_H}     \quad        ,     \quad  \Delta v \sim R^q \; ,
\end{equation}
over a large range of cloud sizes, with   $ 10^{-4}\; - \; 10^{-2}
\;  pc \;   \leq     R   \leq 100\;  pc, \;$
\begin{equation}\label{expos}
1.4    \leq   d_H    \leq   2     ,   \;     0.3  \leq     q  \leq
0.6 \; . 
\end{equation}
These {\bf scaling}  relations indicate a hierarchical structure for the 
molecular clouds which is independent of the scale over the above 
cited range; above $100$ pc in size, corresponding to giant molecular clouds,
larger structures will be destroyed by galactic shear.

These relations appear to be {\bf universal}, the exponents 
$d_H , \; q$ are almost constant over all scales of the Galaxy, and
whatever be  
the observed molecule or element. These properties of interstellar cold 
gas are supported first at all from observations (and for many different 
tracers of cloud structures: dark globules using $^{13}$CO, since the
more abundant isotopic species $^{12}$CO is highly optically thick, 
dark cloud cores using $HCN$ or $CS$ as density tracers,
 giant molecular clouds using $^{12}$CO, HI to trace more diffuse gas, 
and even cold dust emission in the far-infrared).
Nearby molecular clouds are observed to be fragmented and 
self-similar in projection over a range of scales and densities of 
at least $10^4$, and perhaps up to $10^6$.

The physical origin as well as the interpretation of the scaling relations 
 (\ref{vobser}) have been the subject of many proposals. It is not our
 aim here to account for all the proposed models  
of the ISM and we refer the reader to refs.\cite{obser} for a review.

The physics of the ISM is complex, especially when we consider the violent
perturbations brought by star formation. Energy is then poured into 
the ISM either mechanically through supernovae explosions, stellar winds,
bipolar gas flows, etc.. or radiatively through star light, heating or
ionizing the medium, directly or through heated dust. Relative velocities
between the various fragments of the ISM exceed their internal thermal
speeds, shock fronts develop and are highly dissipative; radiative cooling
is very efficient, so that globally the ISM might be considered 
isothermal on large-scales. 
Whatever the diversity of the processes, the universality of the
scaling relations suggests a common mechanism underlying the physics.

  We proposed that self-gravity is the main force at the origin of the 
structures, that can be perturbed locally by heating sources\cite{natu,prd}. 
Observations are compatible with virialised structures at all scales.
 Moreover, it has been suggested that the molecular clouds ensemble is
in isothermal equilibrium with the cosmic background radiation at $T \sim 3 K$
in the outer parts of galaxies, devoid of any star and heating
sources. This colder isothermal medium might represent the ideal
frame to understand the role of self-gravity in shaping the hierarchical
structures. 

\bigskip

In order to compare the properties of the self-gravitating gas with
the ISM it is convenient to express $ m, \; T $ and $ L $ in $ \eta $
in appropriate units. We find from eq.(\ref{defeta})
$$
\eta = 0.52193 \;  { m \; {\cal M}_{\odot}  \over L \; T} \; ,
$$
where $m$ is in multiples of the hydrogen atom mass, $T$ in Kelvin, $L$
in parsecs and  $ {\cal M}_{\odot} $ is the mass of the cloud in units
of solar masses. Notice that $ L $ is many times ($ \sim 10 $) the
size of the cloud. 

The observed parameters of the ISM clouds\cite{obser} yield an $ \eta
$ around $ \sim 2.0 $ for clouds not too large: $ {\cal M}_{\odot} 
< 1000 $. 
Such $ \eta $ is in the range where the
self-gravitating gas exhibits scaling behaviour. 

We conclude that the self-gravitating gas in thermal equilibrium well
describe the observed fractal structures and the scaling relations in
the ISM clouds [see, for example fig. \ref{fig12} and table 2]. Hence,
self-gravity  accounts for the structures in the ISM.

\section{Discussion and Conclusions}

We have presented here a set of new results for the self-gravitating thermal
gas obtained by Monte Carlo and analytic methods. They provide a
complete picture for the thermal self-gravitating gas.

Contrary to the usual hydrostatic treatments \cite{chandra,sas}, we {\bf do not
assume} here an equation of state but we {\bf obtain} the equation of state 
from the partition function [see eq.(\ref{pVnT})]. We find at the same time
that the relevant   variable is here $ \eta^R = G m^2 N/[V^{1/3}  T] $.
The relevance of the ratio $ G m^2 /[V^{1/3}  T] $ has been noticed on
dimensionality  grounds \cite{sas}. However, dimensionality arguments alone
cannot single out the crucial factor $ N $ in the variable $ \eta^R $.

The crucial point is that the thermodynamic limit exist if we
let  $ N \to \infty $ and $ V \to \infty $ {\bf keeping $ \eta^R $ fixed}. 
Notice that $ \eta $ 
contains the ratio $ N \; V^{-1/3} $ and not $ N / V $. This means that
in this thermodynamic limit $ V $ grows as $ N^3 $ and thus the volume density
$ \rho = N/ V $ decreases as $ \sim N^{-2} $. $ \eta $ is to be kept fixed
for a  thermodynamic limit to exist in the same way as the temperature.
 $ p V $, the energy $E$, the free energy, the entropy are functions of 
$ \eta $ and $ T $ times $N$. The chemical potential,
specific heat, etc. are just functions of $ \eta $ and $ T $.

Starting from the partition function of the self-gravitating gas, we
have proved from a microscopic calculation that the local equation
of state $ p(\vec r) = T \; \rho_V(\vec r) $ and the hydrostatic
description are exact. Indeed, the dilute nature of the
thermodynamic limit ($N \sim L \to \infty $ with $N/L$ fixed) together
with the long range nature of the gravitational forces
play a crucial role in the obtention of such ideal gas equation of
state.

We find collapse phase transitions both in the canonical and in the
microcanonical ensembles. They take place at different values of the
thermodynamic variables and are of different nature. In the CE the
pressure becomes large and negative in the collapsed phase.
The  phase transition in the MCE is
sometimes called `gravothermal catastrophe'. We find that the
temperature and pressure increase discontinuously at the MCE transition.
Both are zeroth order phase transitions (the Gibbs free energy is
discontinuous). The two phases cannot coexist in equilibrium since
the pressure has different values at each phase. 

The parameter $ \eta^R $ [introduced in eq.(\ref{defeta})] can be
related to the Jeans length of the system  
\begin{equation}\label{longJ}
d_J = \sqrt{3T \over m} { 1 \over \sqrt{G \, m \, \rho}} \; ,
\end{equation}
where $ \rho \equiv N/V $ stands for the number volume density. Combining
eqs.(\ref{defeta}) and (\ref{longJ}) yields
$$
\eta^R = 3 \left(L \over d_J \right)^2 \; .
$$
We see that the  phase transition  in the canonical ensemble takes
place for $ d_J \sim L $. [The precise numerical value of the
proportionality coefficient depends on the geometry]. For $  d_J > L $
we find the gaseous phase and for $  d_J < L $ the system condenses as
expected. Hence, the collapse phase transition in the canonical
ensemble is related to the Jeans instability.  

The latent heat of the transition ($q$) is {\bf negative} in the CE
transition indicating that the gas releases heat when it collapses
[see eq.(\ref{qsobreT})]. The MCE transition exhibits an opposite
behaviour. The Gibbs free energy increases at the MCE collapse phase
transition (point MC in fig.\ref{fig14}) whereas it decreases at the CE
transition [point T in fig. \ref{fig14}, see eq.(\ref{deltaG})].  Also, the
average distance 
between particles increases at the MCE phase transition whereas it
decreases dramatically in the CE  phase transition. These differences
are related to the MCE constraint keeping the energy fixed whereas in
the CE the system exchanges energy with an external  heat bath keeping
fixed its temperature. The constant energy  constraint in the MCE
keeps the gas stable in a wider domain and makes the collapse
transition softer than in the CE. Notice that the core is much tighter
and the halo much smaller  in the CE than in the MCE [see
figs. \ref{colmc} and  \ref{colc}]. 

\subsection{Hydrodynamical description}

More generally, one can investigate whether a hydrodynamical
description will apply for a self-gravitating gas. One has then to
estimate the mean free path $(l)$ for the particles and compare it
with the relevant scales $ a $ in the system \cite{llkine}. We have,
\begin{equation}\label{clm}
l \sim { 1 \over \rho_V \; \sigma_t } \sim {L^3 \over N \; \sigma_t}
\end{equation}
where $ \rho_V = {N \over L^3} \; \rho $ is the volume density of
particles and $ \sigma_t $ the total transport cross section. 

Due to the
long range nature of the gravitational force, $ \sigma_t $ diverges
logarithmically for small angles. On a finite volume the impact
parameter is bounded by $ L $ and the smaller scattering angle is of
the order of
$$
{ \Delta q \over q} \sim {G\; m^2 \over L \, T}
$$
since $ q = mv \sim \sqrt{m \; T} $ and $ \Delta q \sim {G \, m^2
\over L^2} {L \over v} \sim {G \, m^{5/2} \over L \, \sqrt{T} }$. 

We then have for the transport cross section\cite{llkine},
\begin{equation}\label{sectra}
\sigma_t \sim {(G\, m)^2 \over |\vec v - \vec v'|^4}\log{L\, T \over G\,m^2}
\sim \left({L\, N \over \eta}\right)^2 \log{N \over \eta}
\end{equation}
where we used that $ v \sim \sqrt{T \over m} $ and $ |\vec v - \vec
v'| \sim \Delta q / m $. As we see, the collisions with very large
impact parameters ($\sim L$) dominate the cross-section. 

From eqs.(\ref{clm}) and (\ref{sectra}), we find for the mean free
path: 
\begin{equation}\label{knudsen}
l  \sim {L \over N} \; \left({G\,m^2 \over T \, L}\right)^2
{1 \over \log{T \, L\over G\,m^2}} \sim {L \over N^3} {\eta^2 \over
\log{N \over \eta}} \; ,
\end{equation}
where we have here replaced $ \rho_V $ by $ {N \over L^3} $.
A more accurate estimate introduces the factor $ \rho = e^{\phi} $ in
the denominator of $ l $. This factor for spherical symmetry can vary
up to two orders of magnitude [see \cite{II}]
but it does not change essentially the estimate (\ref{knudsen}). 

We see from eq.(\ref{knudsen}) that  in the thermodynamic limit $ l $
becomes extremely small compared with any length $ a = {\cal O} (N^0)
$ that stays fixed for $ N \to \infty $. We find from
eq.(\ref{knudsen}),
$$
{l \over a} \sim {1 \over N^2} {\eta^2 \over \log{N \over \eta}} \; ,
$$
In conclusion, the smallness of the ratio $l/a$ (Knudsen number)
guarantees that the hydrodynamical description for a self-gravitating
fluid becomes exact in the $ N,\; L \to \infty $ limit for all scales
ranging from the order $ L^0 $ till the order $ L $.

It must me noticed that the time between two collisions $ t_{col}
   = l/v \sim l \, \sqrt{m \over T}  $ is different both from the
   relaxation time and from the  crossing time used in the
   literature. In particular, it is well known that\cite{sas,bt}
   $$
{ t_{crossing} \over t_{relaxation} } \sim {8 \over N} \, \log N \; .
$$
This formula does not concern the time $ t_{col} $ between two
   successive collisions. The time  $ t_{col} $ is indeed very short due to the
   small angle behaviour of the gravitational cross section. For
   constant cross sections one finds a very different result for $
   t_{col} $ [see ref. \cite{bt}].

\bigskip

In refs.\cite{I,II} we thoroughly investigate the
physics of the self-gravitating gas in thermal equilibrium. It is
natural to study now the hydrodynamics of the self-gravitating fluid
using $ p(\vec r) = T \; \rho(\vec r) $ as local equation of state. A
first work on this direction is ref.\cite{seme}.

\end{document}